\documentclass[a4paper,11pt]{article}
\pdfoutput=1 

\usepackage{jheppub} 
\usepackage{braket}
\usepackage[T1]{fontenc} 
\usepackage{amsmath}
\usepackage{dsfont}
\usepackage{subcaption}

\newcommand{\ud}{\,\mathrm{d}}

\title{\boldmath Two intervals R\'enyi entanglement entropy of compact free boson on torus}

\author{Feihu Liu, Xiao Liu}

\affiliation{University of Electronic Science and Technology of China, Chengdu, China}

\emailAdd{liufeihu@std.uestc.edu.cn}
\emailAdd{xiao.liu.tau@gmail.com}

\abstract{We compute the $N=2$ R\'enyi entanglement entropy of two intervals at equal time in a circle, for the theory of a 2d compact complex free scalar at finite temperature. This is carried out by performing functional integral on a genus 3 ramified cover of the torus, wherein the quantum part of the integral is captured by the four point function of twist fields on the worldsheet torus, and the classical piece is given by summing over winding modes of the genus 3 surface onto the target space torus. The final result is given in terms of a product of theta function and certain multi-dimensional theta function. We demonstrate the T-duality invariance of the result. We also study its low temperature limit. In the case in which the size of the intervals and of their separation are much smaller than the whole system, our result is in exact agreement with the known result for two intervals on an infinite system at zero temperature \cite{eeoftwo}. In the case in which the 
separation between the two intervals is much smaller than the interval length, the leading thermal corrections take the same universal form as proposed in \cite{Cardy:2014jwa,Chen:2015cna} for R\'enyi entanglement entropy of a single interval.}

\begin{document}
 
\maketitle
\flushbottom

\section{Introduction}
\label{sec:intro}
Entanglement entropy is one the most unique quantity in quantum field theory and quantum many body systems. It is defined as the von Neumann entropy of the reduced density matrix of the subsystem. 
For dimension $D=2$, one can use the tools of CFT, see \cite{ee-and-cft} for a review. Essentially, to obtain the entanglement entropy is equivalent to calculate the partition function of certain CFTs on the higher genus Riemann surface. It is also known as the replica trick \cite{Holzhey:1994we}:
 Consider now the system is in a pure state with density matrix $\rho=\ket{\Phi}\bra{\Phi}$. For the subsystem $A$ consisted of one or more intervals, the reduced density matrix of is defined by $\rho_A=\mathrm{Tr}_B(\rho)$, where $B$ is the complement of $A$.
 Then one can define the $N$-th R\'enyi entropy
 \begin{equation}
 S_N=\frac{1}{1-N}\ln \mathrm{Tr}\rho_A^N
 \end{equation}
The entanglement entropy is easily obtained by taking the limit
 \begin{equation}
 S_{EE}=\lim_{N\rightarrow 1}S_N.
 \end{equation}
 
 From the path integral point of view, the calculation of $\mathrm{Tr}\rho_A^N$ is equivalent to find the 
 partition function on a $N$-sheet Riemann surface which glued together along $B$ but leave $A$ cut open \cite{ee-and-cft}
 \begin{equation}
 \mathrm{Tr}\rho_A^N=\frac{Z_N(A)}{Z^N}, 
 \end{equation}
 where $Z_N(A)$ is the partition function on the $N$-sheeted Riemann surface and the normalization factor $Z$ is just the original partition function
  $$Z=\mathrm{Tr}e^{-\beta H}.$$
 
In the simplest context of non-compact free boson, the convenient way to calculate $\mathrm{Tr}\rho_A^N$ is by employing the twist fields, which reduce the problem of how to find the higher genus partition function to the problem of calculation
the correlation function of the twist fields \cite{ee-and-cft}
\begin{equation} \label{eq:twistorproduct}
	\mathrm{Tr}\rho_A^N=\prod_{k=0}^{N-1}\braket{\sigma_k(u_1,\bar u_1)\sigma_{-k}(v_1,\bar{v}_1)\cdots},
\end{equation} 
where $k=\{0,1,\cdots,N-1\}$.
Noted that the twist field $\sigma_k$ and the anti-twist field $\sigma_{-k}$ always appear in pairs to create the 
correct branch cuts. Buy using the twist fields, the complexity of the world sheet is transferred to the target 
orbifold space, and one can used the technic from orbifold theory \cite{cft-of-orbifold,multiloop} to calculate the 
correlators of the twist fields.
For now, the entanglement entropy of two disjoint intervals on an infinite system at zero temperature has been studied in \cite{eeoftwo,eeoftwo2}, and one interval on a circle at finite 
temperature has been studied in \cite{Chen:2015cna,1311.1218}.

In this paper, we consider the compactified complex free bosonon on a circle at finite temperature, and study the $N=2$ R\'enyi entanglement entropy of two disjoint intervals.
We should remark that, in case of compactified boson, one should not use \eqref{eq:twistorproduct} directly,
since the different $k$-modes are actually correlated because of the compactification condition. Thus we will take the strategy used in \cite{cft-of-orbifold}: We separate the fundamental field into a classical part and a quantum part, and require that only the classical part see the winding. Therefore one 
can safely regard \eqref{eq:twistorproduct} as the quantum part, in that case we can 
borrow the results in \cite{multiloop} to get the quantum part. 

The classical part are obtained by summing over the independent winding modes. Since for different $k$, the winding numbers are correlated, as a result, the independent wining numbers need to be summed are actually fewer than we have defined. For $N=2$ and two intervals, there are only six independent integers. To be noticed, our method is a little different from \cite{eeoftwo}, in which there are zero modes because of double counting, but after eliminating the zero mode divergence, they are actually the same.
The summation can be expressed as two Riemann-Siegel theta functions, where for each one we have defined a $3\times3$ matrix (see equation \eqref{eq:matrix}). We have to say, it is not obvious at all that the two matrix are strictly positive definite and the have a relation $\Gamma^{-1}=4\Gamma^\prime$, so that we can write the results as Riemann-Siegel theta function.
These properties of the two matrix are highly non-trivial and they indeed represent the T-duality. 

We further study the low temperature expansion. In order to check the consistency with other result, we consider the large system limit, i.e., the subsystem is much smaller than other scale of the system, and we find that the leading term is agreed with the R\'enyi entanglement entropy of two disjoint intervals in an infinite system at zero temperature \cite{eeoftwo}. Further, by considering the separation is much smaller than the length of two intervals, we show that the leading thermal correction is also in consistent with the result in \cite{Chen:2015cna,Cardy:2014jwa}.

The organization of the paper is as follows. In section \ref{sc:review}, we briefly review the method how to calculate correlation function of twist fields. 
 In section \ref{sc:special}, we derive the $N=2$ R\'enyi entropy. In section \ref{sc:lowt}, we obtain the low temperature expansion and compare it with the results. Finally, in section \ref{sc:conclusion} we give our conclusions.

\section{Conformal field theory of orbifold and twist fields}\label{sc:review}

Following \cite{multiloop}, we consider a free compactified complex boson living on 
a Euclidean torus
\begin{equation}
\mathcal{L}=\frac{1}{8\pi}\int \ud z\ud \bar{z}(\partial X \bar{\partial} \bar X+\bar \partial X \partial \bar X), \quad X(z+\pi p+\mathrm{i}\beta\pi q)=
{X}+2\pi R(m+\mathrm{i}n)
\end{equation}
where $p,q$ and $m,n$ are integers, we also have set $\alpha^\prime=2$ in the convention of string theory \cite{joe}. For simplicity, we assume the two compact radii are equal length. Suppose that there are two disjoint intervals on the real axis, 
the replica method is essentially doing the path integral on a $N$-sheeted Riemann surface which are glued together along the two intervals. If in each sheet labeled by $i$ there lives a replica field $\tilde{X}^i$, the gluing simply means that there is a symmetry among these replica fields
\begin{equation}
{X}^i(ze^{2\pi \mathrm{i}},\bar ze^{-2\pi \mathrm{i}})={X}^{i-1}(z,\bar z),
\end{equation}
where we have assumed that $z=0$ is an end point of the intervals.
 After a redefinition of the replica fields \cite{eeoftwo}
\begin{equation}
\tilde{X}^i=\sum_{j=1}^{N}e^{2\pi \mathrm{i}\frac{k_i}{N}j}\tilde{X}^j, \, 0\le k_i<N,
\end{equation}
the new fields $\tilde{X}^i$ satisfy the monodromy condition
\begin{equation}
\tilde{X}^i(ze^{2\pi \mathrm{i}},\bar ze^{-2\pi \mathrm{i}})=e^{2\pi \mathrm{i}\frac{k_i}{N}}\tilde{X}^{i}(z,\bar z).
\end{equation}
Note that the new field $\tilde{X}^i$ indeed lives on the 
original worldsheet torus with the presence of twist fields, while the field $X^j$, which will also be referred to as the replica field, lives on 
the covering genus $3$ Riemann surface.
	
As suggested in \cite{ee-and-cft}, this configuration is equivalent to put twist/antiwist pairs  on the original worldsheet at the ends of the intervals. For example, if we have two intervals labeled by $[z_1,z_2]$ and $[z_3,z_4]$, then there are four insertions $\{\sigma_k(z_1),\sigma_{-k}(z_2),\sigma_k(z_3),\sigma_{-k}(z_4)\}$. We also have the OPEs known as local monodromy condition given by \cite{cft-of-orbifold}
\begin{equation}
  \label{eq:localtwistoperator}
  \begin{cases}
    \partial_z \tilde{X}(z,\bar z) \sigma_k (\omega, \bar \omega) &\sim (z-\omega)^{-(1-\frac{k}{N})}\tau_k  (\omega, \bar \omega), \\
    \partial_z \bar{\tilde{X}}(z,\bar z) \sigma_k (\omega, \bar \omega) &\sim (z-\omega)^{-\frac{k}{N}}\tau_k ^\prime (\omega, \bar \omega), \\
  \partial_{\bar z} \tilde{X}(z,\bar z) \sigma_k (\omega, \bar \omega) &\sim (\bar z-\bar \omega)^{-\frac{k}{N}}\tilde \tau_k  (\omega, \bar \omega),\\
\partial_{\bar z } \bar{\tilde{X}} (z,\bar z) \sigma_k (\omega, \bar \omega) &\sim (\bar z-\bar \omega)^{-(1-\frac{k}{N})}\tilde{\tau}_k ^\prime (\omega, \bar \omega),
\end{cases}
\end{equation}
where $k=\{0,\cdots, N-1\}$. 

Because the insertions of twist fields, the net-twist-zero loops surrounding different subsets of insertions may not be equivalent. Actually, in the most general cases, the number of independent closed loops is $L-2+2g$, where $L$ is the number of twist fields and $g$ is genus of the Riemann surface \cite{multiloop}. The shifts of $X$ along each loop give the global monodromy condition
\begin{equation}
  \label{eq:globalmonodromy-condition}
  \Delta_{\gamma_a} \tilde{X} \equiv \oint_{\gamma_a} \ud z \partial_z \tilde{X} + \oint_{\gamma_a} \ud \bar z \partial_{\bar z }\tilde{X} =v_a,
\end{equation}
where $\gamma_a$ label the closed loops and $\nu_a$ are the shifts which encode the winding number. 


\subsection{Quantum part of the correlation function}

It is convenient to seperate $\tilde{X}$ into a classical part and a quantum part
$$\tilde{X}=\tilde{X}_{qu}+\tilde{X}_{cl},$$ requiring that only the classical part can see the winding
\begin{equation}
\label{eq:quantumpart2}
\begin{split}
\Delta_\gamma \tilde{X}_{qu} &\equiv \oint dz \partial_z \tilde{X}_{qu}(z,\bar z)+\oint d\bar z \partial_{\bar z} \tilde{X}_{qu}(z,\bar z)=0,\\
\Delta_\gamma \tilde{X}_{cl} &\equiv \oint dz \partial_z \tilde{X}_{cl}(z,\bar z)+\oint d\bar z \partial_{\bar z} \tilde{X}_{cl}(z,\bar z)=v.
\end{split}
\end{equation}
We first calculate the quantum part by inserting a stress tensor in the correlation function.
It is known that the twist fields is primary and its OPEs with stress tensor is given by 
\begin{equation}
  \label{eq:OPE}
  T(z)\sigma_i(\omega)\sim \frac{h_i}{(z-\omega)^2}\sigma_i+\frac{1}{z-\omega}\partial_\omega\sigma_i.
\end{equation}
From the Ward identity, one can derive a differential equation of $Z_{qu}$
  \begin{equation}
  \label{eq:pdeofquantumpart}
  \partial_{z_i}\ln Z_{qu}=\lim_{z\rightarrow z_i}\left[ (z-z_i)\braket{\braket{T(z)}}-\frac{h_i}{z-z_i} \right].
\end{equation}
where $h_i=\frac{1}{2}\frac{k_i}{N}(1-\frac{k_i}{N})$ is the conformal dimension of the twist operator and $\braket{\braket{T(z)}}$ is the one point function of stress tensor in the presence of twist fields.

So the main problem is how to construct $T(z)$.  We start with the Green's function:
\begin{equation}
  \label{eq:greenfunction-definition}
  g(z, \omega; z_i) \equiv \frac{\braket{-\partial_z \tilde{X} \partial_\omega \bar{\tilde{X}}\prod \sigma_i(z_i)}}{\braket{ \prod \sigma_i(z_i)}}.
\end{equation}
Taking into account the global monodromy condition \eqref{eq:globalmonodromy-condition}, one should introduce another auxiliary Green's function
\begin{equation}
  \label{eq:auxiliary-green}
  h(\bar z, \omega;z_i)\equiv \frac{\braket{-\partial_{\bar z} \tilde{X} \partial_\omega \bar{\tilde{X}}\prod \sigma_i(z_i)}}{\braket{ \prod \sigma_i(z_i)}},
\end{equation}
which is non-singular as $w\rightarrow z$.
Then $\braket{T(z;z_i)}$ can be obtained by taking a limit
\begin{equation}
  \label{eq:Tz}
  \braket{\braket{T(z;z_i)}}= \lim_{\omega\rightarrow z}\left [ g(z,\omega;z_i)-\frac{1}{(z-\omega)^2} \right ].
\end{equation}

These Green's functions can be constructed by the so called cut abelian differentials\cite{multiloop}. On the torus, it is enough to use the local monodromy and the double period condition to construct the basis of cut abelian differentials. The local monodromy 
is given by
\begin{equation}
  \label{eq:GreenFunction-monodromy}
   g(z,\omega;z_i )\propto
  \begin{cases}
      (z-\omega)^{-2}  & \text{if } z\rightarrow \omega, \\
       (z-z_i )^{-(1-k_i/N )} &  \text{if } z \rightarrow z_i,\\
       (\omega - z_i)^{-k_i/N} & \text{if } \omega, \rightarrow z_i.
  \end{cases}                
\end{equation}
and
\begin{equation}
  \label{eq:GreenFunction-monodromy2}
   h(\bar z,\omega;z_i )\propto
  \begin{cases}
       (\bar z- \bar z_i )^{-k_i/N} &  \text{if } \bar z \rightarrow \bar z_i,\\
       (\omega - z_i)^{-k_i/N} & \text{if } \omega, \rightarrow z_i.
  \end{cases}                
\end{equation}
In case of two pairs of twist/antitwist insertions on the torus, such as $\{\sigma_k(z_1),\sigma_{-k}(z_2),\sigma_k(z_3),\sigma_{-k}(z_4)\}$, one can define the four cut abelian differentials:
\begin{equation}
  \label{eq:abeliandiff}
  \begin{split}
  w^1(z)=\prod_{i=1}^4\vartheta_1(z-z_i)^{-(1-\frac{k_i}{N})}\vartheta_1(z-z_{\alpha_1}-Y_1)\vartheta_1(z-z_{\alpha_2}),\\
 w^2(z)=\prod_{i=1}^4\vartheta_1(z-z_i)^{-(1-\frac{k_i}{N})}\vartheta_1(z-z_{\alpha_2}-Y_1)\vartheta_1(z-z_{\alpha_1}),\\
 w^3(z)=\prod_{i=1}^4\vartheta_1(z-z_i)^{-\frac{k_i}{N}}\vartheta_1(z-z_{\beta_1}-Y_2)\vartheta_1(z-z_{\beta_2}),\\
 w^4(z)=\prod_{i=1}^4\vartheta_1(z-z_i)^{-\frac{k_i}{N}}\vartheta_1(z-z_{\beta_2}-Y_2)\vartheta_1(z-z_{\beta_1}).
 \end{split}
\end{equation}
where $0\le k_i<N$ are integers. $Y_1$ and $Y_2$ can be determined by requiring that $w^i(z)$ are doubly periodic. Their values are given by 
\begin{equation}
Y_1=\sum_{i=1}^4(1-\frac{k_i}{N})z_i-\sum_{i=1}^{2}z_{\alpha_i}, \quad Y_2=\sum_{i=1}^4\frac{k_i}{N}z_i-\sum_{i=1}^{2}z_{\beta_i}.
\end{equation}
The points $\{ z_{\alpha_1},z_{\alpha_2} \}$ is a subset of the four twist insertions. To make the first two cut differentials 
linear independent, one need to be careful not to choose the subset in which $Y_1=0$. This is the only constrains for 
choosing $\{z_{\alpha_1},z_{\alpha_2}\}$. Indeed the first two functions span the space of cut differentials \cite{multiloop}, which can be used to construct $\braket{\partial_z \tilde{X}}$ and $\braket{\partial_{\bar z} {\bar{ \tilde{X}}}}$. For the similar reason, $\braket{\partial_{\bar z} \tilde{X}}$ and $\braket{\partial_z {\bar {\tilde{X}}}}$ can be represented by the linear combinations of   ${\bar w}^3(\bar z)$ and ${\bar w}^4(\bar z)$.
Generally, there are no constraints of how to choose $k_i$s. However, in order to create the correct branch cuts, one should fix $\{k_1,k_2,k_3,k_4 \}$ to be $\{k,N-k,k,N-k \}$ for any $0\le k<N$, i.e., the twist and antitwist fields should appear in pairs.

\begin{figure}[h]
  \centering
  \includegraphics[width=4in]{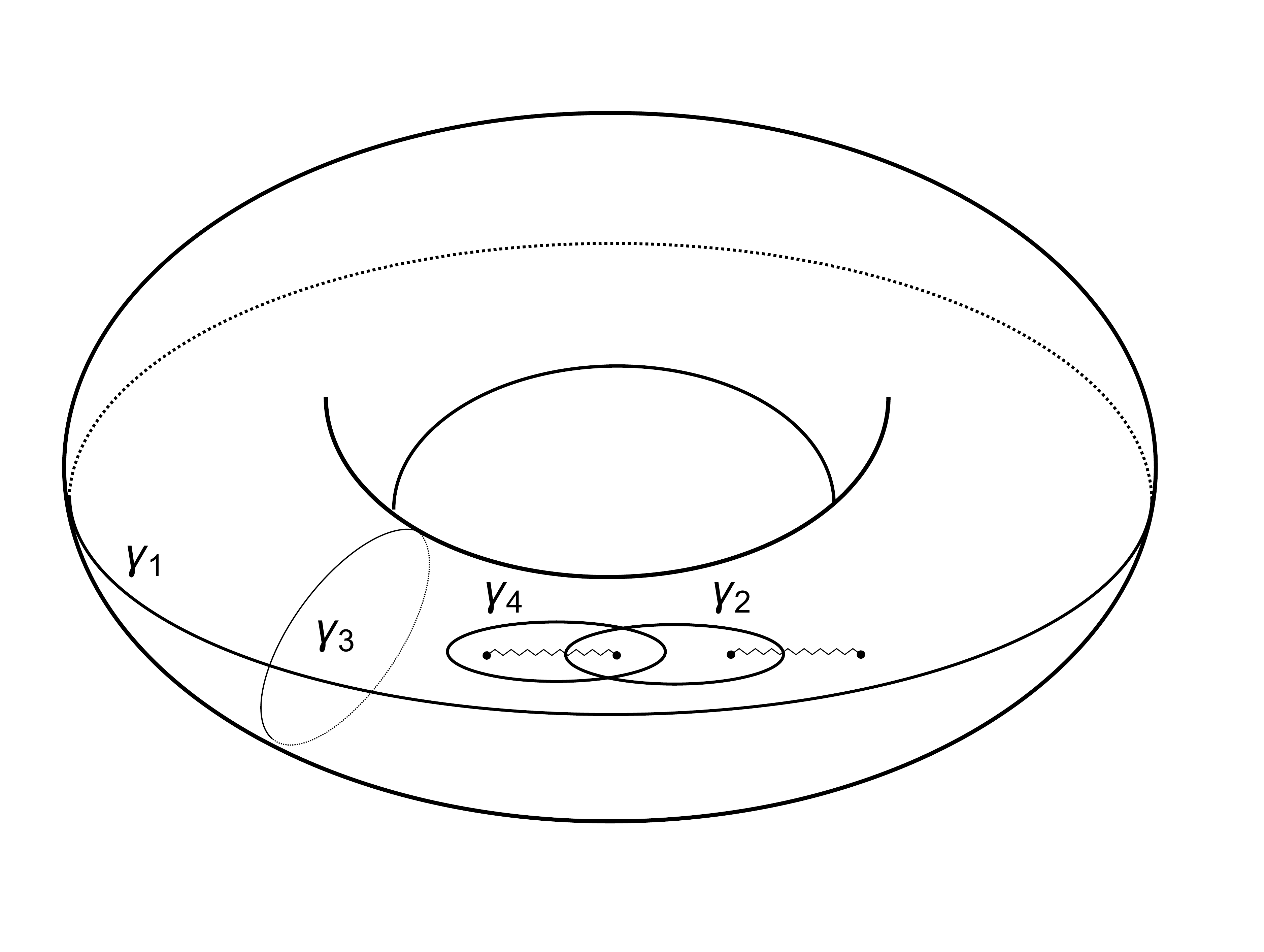}
  \caption{\label{fig:loops} Four independent closed loops.}
\end{figure}

By using the cut abelian differentials \eqref{eq:abeliandiff}, one can fix the Green's function up to some non-singular functions:
\begin{equation}
\label{eq:greenfunctionsolution_undetermined}
\begin{split}
g(z,\omega)&= g_s(z,\omega)-\sum_{i=1}^{2}A_{ij}w^j(\omega) w^i(z),\\
h(\bar z,\omega)&= -\sum_{j=3}^4 B_{ji}w^i(\omega) \bar{w}^j(\bar z).
\end{split}
\end{equation}
The four function $A_{ij}w^j(\omega),B_{ji}w^i(\omega)$ can be determined up to normalization by imposing the global monodromy conditions
\begin{equation}\label{eq:globalqu}
\oint_{\gamma_a} dz g(z,\omega)+\oint_{\gamma_a} d \bar z h(\bar z,\omega)=0.
\end{equation} 
These equations \eqref{eq:globalqu} can be solved by introducing the cut period matrix
$W_a^i$ defined by
\begin{equation}
  \label{eq:cutperiodmatrix}
  \begin{split}
  W_a^i &\equiv \oint _{\gamma_a} dz w^i(z), \,i=1,2\\
  W_a^j &\equiv \oint_{\gamma_a} d\bar z \bar{w}^j( \bar z),\,j=3,4,
  \end{split}
\end{equation}
where $\gamma_{a}$ represent the independent closed loops. In this paper, we consider four twist insertions on the worldsheet torus. Thus there are four independent net-twist-zero loops, we can chosen them as described in figure \ref{fig:loops}.

After solving the equations \eqref{eq:globalqu}, the Green's function can be written as
\begin{equation}
  \label{eq:greenfunctionsolution}
  \begin{split}
  g(z,\omega)&= g_s(z,\omega)-\sum_{i=1}^{2}w_i(z)\sum_{a=1}^4(W^{-1})_i^a\oint_{\gamma_a}dy g_s(y,\omega),\\
  h(\bar z,\omega)&= -\sum_{j=3}^4\bar{w}_j(\bar z)\sum_{a=1}^{4} (W^{-1}) _j^a \oint_{\gamma_a}dy g_s(y,\omega),
  \end{split}
\end{equation}
where the singular part $g_s(z,w)$ is given by \cite{multiloop}
\begin{equation}
\label{eq:gs}
	 g_s(z,\omega)=\prod_{i=1}^4\vartheta_1(z-z_i)^{-(1-\frac{k_i}{N})}\prod_{i=1}^4\vartheta_1(w-z_i)^{-\frac{k_i}{N}}\left[\frac{\vartheta_1^\prime (0)}{\vartheta_1(z-\omega)}\right]^2 P(z,\omega).
\end{equation}
However the exact form of $P(z,\omega)$ turns out to be irrelevant in the end.

In general, the period matrix $W$ is a $4\times 4$ matrix. As long as the two twist fields don't coincide, $W$ should be non-degenerate.
Therefore, by using equation \eqref{eq:pdeofquantumpart} and integrating \eqref{eq:Tz}, one can get the quantum part \cite{multiloop}
\begin{equation}
  \label{eq:quantumpart}
  \begin{split}
    Z_{qu}(k,N)=&f(\tau,k)|\det{W}|^{-1}\vartheta_1(Y_1) \bar{\vartheta}_1(Y_2)\vartheta_{34}\bar{\vartheta}_{12}\\
    &\left[ \vartheta_{12}\vartheta_{14}\vartheta_{23}\vartheta_{34} \right]^{-\frac{k}{N}(1-\frac{k}{N})} (\vartheta_{13})^{-(1-\frac{k}{N})(1-\frac{k}{N})}(\vartheta_{24})^{-\frac{k}{N}\frac{k}{N}}\\
    &\left[ \bar \vartheta_{12} \bar \vartheta_{14} \bar \vartheta_{23} \bar \vartheta_{34} \right]^{-\frac{k}{N}(1-\frac{k}{N})}(\bar \vartheta_{24})^{-(1-\frac{k}{N})(1-\frac{k}{N})}(\bar{\vartheta}_{13})^{-\frac{k}{N}\frac{k}{N}},
  \end{split}
\end{equation}
where we have denote $\vartheta_1(z_i-z_j)$ by $\vartheta_{ij}$. Note that $f(\tau,k)$
is an unfixed function came from the integration of the differential equation.

\subsection{Classical part of the correlation function}
The classical contribution can be obtained by finding the normalized classical solution and then substituting back into the action
\begin{equation}
  \label{eq:classicalaction}
  S_{cl}=\frac{1}{16 \pi} \int d^2z (\partial_z\tilde{X}_{cl} \partial_{\bar z}\bar{\tilde{X}}_{cl}+\partial_{\bar z}\tilde{X}_{cl} \partial_{z}\bar{ \tilde{X}}_{cl}).
\end{equation}
The classical solutions can be written as the linear combination of the abelian differentials
\begin{equation}
  \label{eq:linearequation}
  \begin{split}
    \partial_z\tilde{X}_{cl}(z,\bar z)=a_i w^i(z), \, i=1,2, \\
    \partial_{\bar z}\tilde{X}_{cl}(z, \bar z)= b_j  \bar w^j(\bar z), \, j=3,4.
  \end{split}
\end{equation}
Plugging \eqref{eq:linearequation} into the global monodromy condition \eqref{eq:globalmonodromy-condition}, 
we get four linear equations with four unknowns
\begin{equation}
	\label{eq:linerequations}
	\oint_{\gamma_a}dz \partial_z \tilde{X}_{cl}+\oint_{\gamma_a}d\bar{z} \partial_{\bar{z}} \tilde{X}_{cl}=v_a, \quad (a=1,2,3,4).
\end{equation}
The solutions are given by
\begin{equation}
  \label{eq:aandb}
  a_i={(W^{-1})_i}^av_a , \qquad b_j= {(W^{-1})_j}^av_a.
\end{equation}
Then the action can be written as
\begin{equation}
  \label{eq:classicalaction2}
  S_{cl}=\frac{1}{16 \pi} v_a \bar v_b \left[ {(W^{-1})_{i_1}}^a ( {\bar  W^{-1})_{i_2}}^b(w^{i_1}, w^{i_2})+ {(W^{-1})_{j_1} }^a{(\bar W^{-1})_{j_2}}^b(w^{j_1}, w^{j_2})   \right],
\end{equation}
where $ i_1,i_2\in\{1,2\}$ and $ j_1,j_2\in\{3,4\}$. We have also defined the inner product of the cut differentials:
\begin{equation}
  \label{eq:innerproduct2}
  (w^{i}, w^{i}) \equiv i\int_R w^{i} \wedge \bar w^{i}, \quad w^i=w^i(z) dz.
\end{equation}
The inner product can be calculated by using Stokes theorem. A detailed calculation can be found in the appendix \ref{sc:innerproduct}.
The full partition function now is
\begin{equation}
\label{eq:fullpartionfunction}
	Z(N)=\sum_{v,\bar v}\left(\prod_{k=0}^{N-1} Z_{qu}(k,N)e^{-S_{cl}(v,\bar v)}\right)
\end{equation}
Note that the summation has been moved out of the product because the different $k$-modes are correlated.
 
 \section{$N=2$ R\'enyi entropy} 
 In this section we calculate the $N=2$ R\'enyi entanglement entropy of two intervals on a circle at finite temperature, which is the most simple case one can have. Notice that in terms of the replica fields $X^i$, the 
New field $\tilde{X}$ is defined for different $k$ accordingly.
For $k=0$,
 \begin{equation}\label{eq:replica0}
 \tilde{X}^0=X^0+X^1
 \end{equation}
 while for $k=1$
 \begin{equation}\label{eq:replica1}
 \tilde{X}^1=X^0-X^1.
 \end{equation}
 The action for the new field $\tilde{X}$ is changed to
 \begin{equation}
 S(\tilde{X},\bar{\tilde{X}})=\frac{1}{16\pi}\int d^2z (\partial_z\tilde{X}_{cl} \partial_{\bar z}\bar{\tilde{X}}_{cl}+\partial_{\bar z}\tilde{X}_{cl} \partial_{z}\bar{ \tilde{X}}_{cl})
 \end{equation}
We calculate the contribution from different $k$ separately and then multiply them together, after that we sum over the winding number to get the total partition function.

\subsection{$k=0$} \label{sc:special}
For $k=0$, the twist fields are trivial, thus the quantum part can be treated as the 
correlation function of identity operators, which is just the partition function of free boson on the torus \cite{cft}.
\begin{equation}\label{eq:zqu0}
Z_{qu}^0(\tau)=\frac{1}{|\mathrm{Im}\tau||\eta(\tau)|^4}.
\end{equation}
Now we consider the summation of winding mode.
For $k=0$, the classical solution of $\partial X$ is just a double periodic holomorphic function, so it should be a constant
\begin{equation} \label{eq:solution0}
\partial \tilde{X}^0_{cl}=c_1, \, \bar{ \partial} \tilde{X}^0_{cl}=c_2.
\end{equation}
The solutions \eqref{eq:solution0} should be normalized by imposing the global monodromy condition
\begin{equation}
\begin{split}
&\int_{\gamma_1}\partial \tilde{X}^0_{cl} \ud z+ \int_{\gamma_1}\bar \partial \tilde{X}^0_{cl} \ud \bar z=u_1\\
&\int_{\gamma_3}\partial \tilde{X}^0_{cl} \ud z+ \int_{\gamma_3}\bar \partial \tilde{X}^0_{cl} \ud \bar z=u_3,
\end{split}
\end{equation}
where $u_1$ and $u_3$ are denoted by
\begin{equation}
\begin{split}
u_1&=2\pi R(m_0^1+\mathrm{i}n_0^1+m_1^1+\mathrm{i}n_1^1)\\
u_3&=2\pi R(m_0^3+\mathrm{i}n_0^3+m_1^3+\mathrm{i}n_1^3).
\end{split}
\end{equation}
Noted that the superscript $\{1,3\}$ label the different loops, the subscript $\{0,1\}$ label the different replica and $\{m,n\}$ come from the real and imaginary part accordingly.
These equations can be easily solved by
\begin{equation}
c_1=\frac{\mathrm{i} \beta u_1- u_3}{2\mathrm{i}\pi\beta}, \,c_2=\frac{\mathrm{i} \beta u_1+ u_3}{2\mathrm{i}\pi\beta}.
\end{equation}
Then plugging into the classical action \eqref{eq:classicalaction}, one can get the classical contribution
\begin{equation}
\label{eq:third}
\begin{split}
S_{cl}^0(u_1,u_3)&=\frac{|\mathrm{i}\beta u_1-u_3|^2+|\mathrm{i} \beta u_1+ u_3|^2}{32\beta \pi}\\
&=-2 \pi \mathrm{i}\left(  {m^{\prime \prime}}^T\cdot \Xi \cdot m^{\prime \prime}+{n^{\prime \prime}}^T\cdot \Xi \cdot n^{\prime \prime}\right),
\end{split}
\end{equation}
where $m^{\prime \prime}\equiv \{m_0^1,m_1^1,m_0^3,m_1^3\}$, $n^{\prime \prime}\equiv \{n_0^1,n_1^1,n_0^3,n_1^3\}$ and the matrix $\Xi$ is given by
\begin{equation}
\Xi=\frac{ \mathrm{i} R^2}{8\pi^2}
\begin{pmatrix}
\beta & \beta  & 0 & 0  \\
\beta & \beta  & 0 & 0 \\
0 & 0 & \frac{1}{\beta}& \frac{1}{\beta}\\
0 & 0 & \frac{1}{\beta}&\frac{1}{\beta}
\end{pmatrix}.
\end{equation}

\subsection{$k=1$}

\subsubsection{Quantum part for $k=1$}
Let's now consider the case of $k=1$.
We assign the two intervals the same length $x$, the distance between them is given by $y$, and all 
the twist operators lie on the real cycle of the torus, as shown in figure \ref{fig:cut}.
As a consequence, there are only two independent cut abelian differentials
\begin{equation}
\label{eq:cutdifferentials}
\begin{split}
	w^1(z)&= \prod_{i=1}^4\vartheta_1(z-z_i)^{-1/2}\vartheta_1(z-x-\frac{\pi -2x-y}{2}) \vartheta_1(z-x-y-\frac{\pi -2x-y}{2}) \\
	&= \frac{\vartheta_1(z-\frac{\pi}{2}+\frac{y}{2})^{1/2} \vartheta_1(z-\frac{\pi}{2}-\frac{y}{2})^{1/2}}{\vartheta_1(z-\frac{\pi}{2}+x+\frac{y}{2})^{1/2} \vartheta_1(z-\frac{\pi}{2}-x-\frac{y}{2})^{1/2}}=w^3(z),\\
	w^2(z)&= \prod_{i=1}^4\vartheta_1(z-z_i)^{-1/2}\vartheta_1(z-\frac{\pi -2x-y}{2}) \vartheta_1(z-2x-y-\frac{\pi -2x-y}{2})\\
	&= \frac{\vartheta_1(z-\frac{\pi}{2}+x+\frac{y}{2})^{1/2} \vartheta_1(z-\frac{\pi}{2}-x-\frac{y}{2})^{1/2}}{\vartheta_1(z-\frac{\pi}{2}+\frac{y}{2})^{1/2} \vartheta_1(z-\frac{\pi}{2}-\frac{y}{2})^{1/2}}=w^4(z).
\end{split}
\end{equation}
We also define the period matrix in appendix \ref{sc:proof}

The quantum part for $k=1$ can be calculated directly by using \eqref{eq:quantumpart}
\begin{equation}
\label{eq:zqu1}
Z_{qu}^1=f(\beta)\frac{1}{|\det{W}|}\frac{|\vartheta_1(x+y)||\vartheta_1(x)|}{\sqrt{|\vartheta_1(2x+y)||\vartheta_1(y)|}},
\end{equation}
where $f(\beta)$ is an undetermined function came from the integration of $\partial_{z_i}Z_{qu}$. We will fix it later by factorizing the total partition function on the torus partition function of compact free boson .

\subsubsection{Classical part for $k=1$}
Given \eqref{eq:cutdifferentials}, one can expand the cut differential as
\begin{equation*}
\begin{split}
    \partial_z\tilde{X}_{cl}^1(z,\bar z)=a_i\omega^i(z), \, i=1,2 \\
    \partial_{\bar z}\tilde{X}_{cl}^1(z, \bar z)= b_i \bar \omega^i(\bar z),\, i=1,2.
  \end{split}
\end{equation*}
The coefficients can be determined by solving the global monodromy condition
\begin{equation}
  \Delta_\gamma \tilde{X}_{cl} = \oint dz \partial_z \tilde{X}_{cl}(z,\bar z)+\oint d\bar z \partial_{\bar z} \tilde{X}_{cl}(z,\bar z)=v_a.
\end{equation} 
Before solving the equations, we need to be sure that the cut period matrix is 
non-degenerate. This is true as long as the two insertions of twist field don't collide, i.e., $x\ne 0$ and $y\ne 0$, which can be checked numerically. 
Substituting the solutions back into the action, we get
\begin{equation}
\label{eq:classicalpart}
	S_{cl}=\frac{1}{16 \pi} Tr \left[ M\cdot W^{-1}\cdot G \cdot (\bar{W}^{-1})^{ \mathrm{ T }}\right], 
\end{equation}
where $W^{-1}$ and $\bar W^{-1}$ is the inverse of the cut period matrix and its conjugation, 
$M$ is defined by $M_{ab}\equiv v_a \bar v_b$, $G$ is given by the inner product of $\omega^i$
\begin{equation}
\label{eq:G}
	G^{ij}=(\omega^i,\omega^j),
\end{equation}
where $i,j\in\{1,2,3,4\}$.
Since we put all the twist insertions on the real cycle, giving the bilinear relation \ref{eq:innerproduct1}, one can show that $G$ is a block diagonal matrix
\begin{equation}
G=
\begin{pmatrix}
H & 0 \\ 0 & H 
\end{pmatrix},
\end{equation} 
where
\begin{equation}
\begin{split}
& H=\\
&\begin{pmatrix}
2 \mathrm{i} {W_1}^1 {W_3}^1+\mathrm{i} {W_2}^1{W_4}^1 & \mathrm{i} ({W_1}^2 {W_3}^1+{W_1}^1 {W_3}^2)+\frac{\mathrm{i}}{2}({W_2}^2{W_4}^1+{W_2}^1{W_4}^2) \\ 
\mathrm{i} ({W_1}^2 {W_3}^1+{W_1}^1 {W_3}^2)+\frac{\mathrm{i}}{2}({W_2}^2{W_4}^1+{W_2}^1{W_4}^2) & 2 \mathrm{i} {W_1}^2 {W_3}^2+\mathrm{i} {W_2}^2{W_4}^2 
\end{pmatrix}.
\end{split}
\end{equation} 

\begin{figure}[h]
	\centering
	\includegraphics[width=4in]{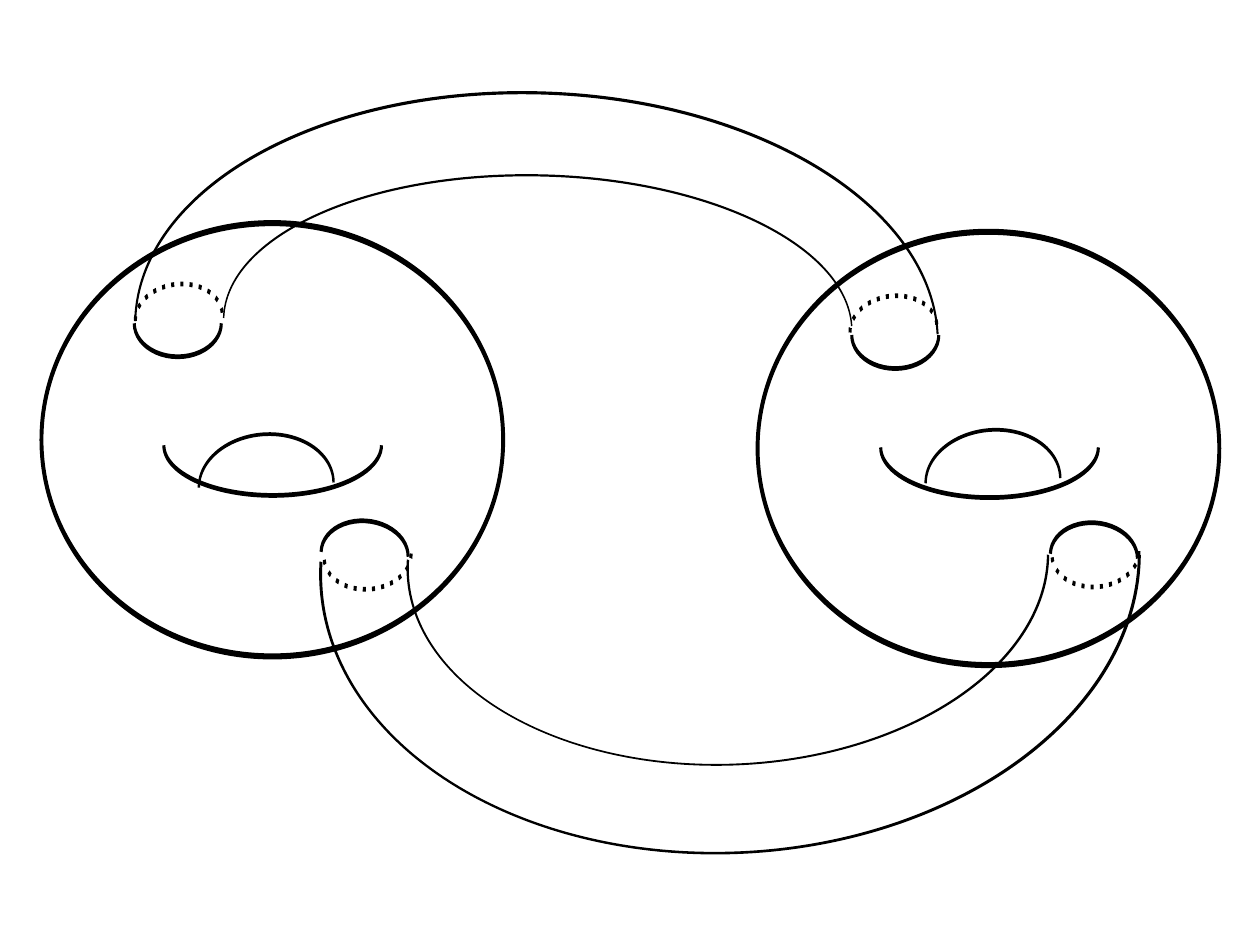}
	\caption{\label{fig:covering} The $N=2$ covering surface with two cuts gluing together.}
\end{figure}

In case of $N=2$, $W_a^i$ are either pure imaginary or real. This feature make it much simpler for lattice summation. We introduce eight arbitrary real functions $\{a,b,c,d,e,f,g,h\}$ and denote the cut period matrix by
\begin{equation}
\label{eq:realimaginary}
{W_a}^i=
\begin{pmatrix}
a & h & a & h  \\
f & g & f & g \\
\mathrm{i} b & \mathrm{i} c & -\mathrm{i} b & -\mathrm{i}c \\
\mathrm{i} d & \mathrm{i} e & -\mathrm{i} d & -\mathrm{i}e
\end{pmatrix}.
\end{equation}
After some algebra, the \eqref{eq:classicalpart} can be divided into two parts, the first half is
 \begin{equation}
 \label{eq:firsthalf}
 S_{cl}^1(v_1,v_2)=\frac{1}{16 \pi}\left( A |v_1|^2+B(v_1\bar v_2+ v_2 \bar v_1 )+C |v_2|^2 \right),
 \end{equation}
where $A,B,C$ are given by
\begin{equation}\label{eq:coefficient}
\begin{split}
A&= \mathrm{i} \frac{-{W_2}^2{W_3}^1+{W_2}^1{W_3}^2}{{W_1}^2{W_2}^1-{W_1}^1{W_2}^2}\\
B&= \mathrm{i} \frac{-2{W_1}^2{W_3}^1+2{W_1}^1{W_3}^2+{W_2}^2{W_4}^1-{W_2}^1{W_4}^2}{-4{W_1}^2{W_2}^1+4{W_1}^1{W_2}^2}\\
C&= \mathrm{i} \frac{-{W_1}^2{W_4}^1+{W_1}^1{W_4}^2}{-2{W_1}^2{W_2}^1+2{W_1}^1{W_2}^2}.
\end{split}
\end{equation}
The parameterization of the shifts $v_a$ are given as follows.
In case of $N=2$, the ramified covering surface is rather simple, see figure \ref{fig:covering}. Therefore we don't need to introduce a complicated target space as did in \cite{ee-and-cft}
\cite{eeoftwo}.
Taking into account the definition \eqref{eq:replica1}, the four shift vectors can be written as
\begin{equation}\label{eq:parameterization}
	\begin{split}
	v_1&=2\pi R\left[(m_0^1-m_1^1)+\mathrm{i}(n_0^1-n_1^1)\right]\\
	v_2&=2\pi R\left[(m_0^2-m_1^2)+\mathrm{i}(n_0^2-n_1^2)\right]\\
		v_3&=2\pi R\left[(m_0^3-m_1^3)+\mathrm{i}(n_0^3-n_1^3)\right]\\
			v_4&=2\pi R\left[(m_0^4-m_1^4)+\mathrm{i}(n_0^4-n_1^4)\right]
	\end{split}
\end{equation}
Notice that the superscript $\{1,2,3,4\}$ label the different loops and the subscript $\{0,1\}$ label the different replica. The $\{m,n\}$ represent the real and imaginary part accordingly.
However, as we mentioned before, the different winding modes of $k=0$ and $k=1$ are actually correlated. In case of $k=0$, the shift vectors corresponding to $\gamma_2$ and $\gamma_4$ are trivial, i.e., equals to zero, this actually is a constraint condition
\begin{equation}\label{eq:contraints}
\begin{split}
m_0^2=-m_1^2 &, \, n_0^2=-n_1^2\\
m_0^4=-m_1^4 &, \, n_0^4=-n_1^4.
\end{split}
\end{equation}
We should impose these constraints into the summation, as a result, the number of independent integers is reduced, which is $12$ rather than $16$. It is reasonable by noticing that the covering surface exactly has genus $g=3$, as show in figure \ref{fig:covering}. Therefore the independent winding modes of a single scalar should be parameterized by $3$ complex vectors, or $6$ real winding numbers. 
Here we have used a different 
 approach comparing to \cite{eeoftwo}, in which they used the orbifold method: the world sheet remains simple but the target space becomes a complicated orbifold. We reported their method in the appendix \ref{sc:oldmethod}.
The only difference between the two methods is that there is a zero mode in the orbifold approach, after absolving the zero mode divergence into the normalization constant, the two methods actually 
agree with each other.
We are going to use the parameterization \eqref{eq:parameterization} and \eqref{eq:contraints}, as we will see, there is no zero mode.

Putting \eqref{eq:firsthalf}, \eqref{eq:parameterization} and \eqref{eq:contraints} all together, we find that 
the first half of the classical action for the field $\tilde{X}^1$ which only depends on $v_1,v_2$, 
is 
\begin{equation}\label{eq:first}
S_{cl}^1(v_1,v_2)=-2 \pi \mathrm{i} ( m^T\cdot \Omega \cdot m+n^T\cdot \Omega \cdot n),
\end{equation}
where $m\equiv\{m_0^1,m_1^1,m_0^2\}\in \mathbb{Z}^3$, $n\equiv\{n_0^1,n_1^1,n_0^2\} \in \mathbb{Z}^3$ and $\Omega$ is a symmetry matrix
\begin{equation}
\Omega=\frac{ \mathrm{i} R^2}{8}
\begin{pmatrix}
A & -A & 2B \\
-A & A & -2B\\
2B & -2B & 4C
\end{pmatrix}.
\end{equation}

The second half of the classical action for the field $\tilde{X}^1$, which only depends on $v_3, v_4$, can be found by the same way, and is given by
\begin{equation}
	\label{eq:second}
	S_{cl}^2(v_3,v_4)=-2 \pi \mathrm{i}\left(  {m^\prime}^T\cdot \Omega^\prime \cdot m^\prime+{n^\prime}^T\cdot \Omega^\prime \cdot n^\prime\right),
\end{equation}
where $m^\prime \equiv\{m_0^3,m_1^3,m_0^4\}\in \mathbb{Z}^3$, $n^\prime \equiv\{n_0^3,n_1^3,n_0^4\}\in \mathbb{Z}^3$ and
\begin{equation}
	\Omega^\prime=\frac{ \mathrm{i} R^2}{ 8}
	\begin{pmatrix}
		A^\prime  & -A^\prime  &2 B^\prime  \\
		-A^\prime & A^\prime  & -2B^\prime \\ 
		2B^\prime  & -2B^\prime &4 C^\prime
	\end{pmatrix}.
\end{equation}
The matrix elements $A^\prime$, $B^\prime$ and $C^\prime$ given below
\begin{equation}\label{eq:coefficient2}
	\begin{split}
		A^\prime&= \mathrm{i} \frac{{W_1}^2{W_4}^1-{W_1}^1{W_4}^2}{-{W_3}^2{W_4}^1+{W_3}^1{W_4}^2}\\
		B^\prime&= \mathrm{i} \frac{-2{W_1}^2{W_3}^1+2{W_1}^1{W_3}^2+{W_2}^2{W_4}^1-{W_2}^1{W_4}^2}{-4{W_3}^2{W_4}^1+4{W_3}^1{W_4}^2}\\
		C^{\prime}&= \mathrm{i} \frac{-{W_2}^2{W_3}^1+{W_2}^1{W_3}^2}{-2{W_3}^2{W_4}^1+2{W_3}^1{W_4}^2}.
	\end{split}
\end{equation}

\subsection{Results}
\subsubsection{Lattice summation}\label{sec:plot}
Substituting all the intermediate results into \eqref{eq:fullpartionfunction}, now we can do the lattice summation 
\begin{equation}
\label{eq:classicalsummation}
Z_{cl}=\sum_{v_1,v_2,v_3,v_4,u_1,u_3} e^{-S_{cl}^0(u_1,u_3)}e^{-S_{cl}^1(v_1,v_2)}e^{-S_{cl}^2(v_3,v_4)}
\end{equation} 
After some algebra, we find
\begin{equation}
\label{eq:instantonsum}
Z_{cl}=\left(\sum_{m \in \mathbb{Z}^3} e^{2 \pi \mathrm{i}  m^T\cdot\frac{ \mathrm{i} R^2}{4} \Gamma \cdot m}\right)^2 \left(\sum_{m^\prime \in \mathbb{Z}^3} e^{2 \pi \mathrm{i}  {m^\prime}^T\cdot \frac{ \mathrm{i} R^2}{4}\Gamma^\prime \cdot m^\prime}\right)^2,
\end{equation}
where 
  \begin{equation}\label{eq:matrix}
  \Gamma=
  \begin{pmatrix}
A+\beta & -A+\beta & 2B  \\
-A+\beta & A+\beta & -2B \\
2B & -2B & 4C
  \end{pmatrix},\,
    \Gamma^\prime=
    \begin{pmatrix}
    A^\prime+\frac{1}{\beta}  & -A^\prime+\frac{1}{\beta}  &2 B^\prime  \\
    -A^\prime+\frac{1}{\beta} & A^\prime+\frac{1}{\beta}  & -2B^\prime \\ 
    2B^\prime  & -2B^\prime &4 C^\prime 
    \end{pmatrix}
  \end{equation}
The matrix $\Gamma$ and $\Gamma^\prime$ are symmetric and real, most importantly, they are positive definite. Although it is hard to prove analytically, it can be easily checked numerically.
Thus, by using Riemann-Siegel theta function, the classical part \eqref{eq:instantonsum} can be written as
\begin{equation}
\label{eq:classical}
	Z_{cl}=\Theta(0|\frac{\mathrm{i}R^2}{4} \Gamma)^2 \Theta(0|\frac{\mathrm{i}R^2}{4} \Gamma^\prime)^2.
\end{equation}
It is worth to mention that the dimension of $\Gamma$ and $\Gamma^\prime$ is $3$, which is exactly 
the genus of the $N=2$ ramified covering surface of the torus with two cuts, see figure \ref{fig:covering}.

Given the equations \eqref{eq:zqu0},\eqref{eq:zqu1} and \eqref{eq:classical}, one 
can obtain the total partition function
\begin{equation}\label{eq:z}
\begin{split}
Z&=Z_{qu}^{0}Z_{qu}^{1}Z_{cl}\\
&=\frac{f(\beta)}{\beta|\eta(\mathrm{i}\beta)|^4}\frac{1}{|\det{W}|}\frac{|\vartheta_1(x+y)||\vartheta_1(x)|}{\sqrt{|\vartheta_1(2x+y)||\vartheta_1(y)|}}\Theta(0|\frac{\mathrm{i}R^2}{4} \Gamma)^2 \Theta(0|\frac{\mathrm{i}R^2}{4} \Gamma^\prime)^2,
\end{split}	
\end{equation}
where $f(\beta)$ need to be fixed. This can be done by analyzing the behavior of $Z$ in the limit
of $x\rightarrow 0$. 

We already know that the conformal dimension of the twist field for $N=2$ is $(1/8,1/8)$. As one pair of twist and antitwist fields come together, they should factor onto 
the identity operator according to the OPE:
\begin{equation}
\label{eq:opeoftwistfields}
\sigma_{1/2}(z_1,\bar{z}_1)\sigma_{1/2}(z_2,\bar{z}_2)\sim (z_1-z_2)^{-1/4}(\bar{z}_1-\bar{z}_2)^{-1/4}\mathrm{1}(z_z,\bar{z}_2).
\end{equation}
However it is not clear that, the complete genus $3$ partition function should behave the same way at $x \rightarrow 0$ as described by the above OPE. 
Here we assume that in the small $x$ limit, the leading singular behavior for the partition function nonetheless coincides with that of the OPE, and fix $f(\beta)$ by demanding 
\begin{equation}
\lim_{x\rightarrow 0}Z\sim \frac{Z_{b}^2}{x},
\end{equation}
where $Z_b$ is the partition function of the compact complex free scalar on the torus given by \cite{cft}
\begin{equation}\label{eq:zb}
\begin{split}
Z_b&=\frac{R^2}{2}\frac{1}{\mathrm{Im}(\tau)|\eta(\tau)|^4}\left[\sum_{m,m^\prime}\exp\left(-\frac{\pi R^2|m\tau-m^\prime|^2}{2 \mathrm{Im}(\tau)}\right)\right]^2\\
&=\frac{1}{|\eta(\tau)|^4}\vartheta_3(0|\frac{\mathrm{i}\beta R^2}{2})^2\vartheta_3(0|\frac{\mathrm{i}2\beta}{ R^2})^2,
\end{split}
\end{equation}
where in the last line of \eqref{eq:zb} we have assumed $\tau=\mathrm{i}\beta$ is pure imaginary and resummed over $m^\prime$.

To get the leading singular term, we start by expanding the elements of period matrix with respect to $x$, the general form of the contour integrals can be expressed by
\begin{equation}\label{eq:pinchedmatrix}
\begin{split}
	{W_1}^1={W_1}^2&= \pi +O(x^2),\\
	{W_2}^1= -2y+2x(F(y,\beta)&+G(y,\beta)\log(x)) +O(x^2),\\
	{W_2}^2= -2y-2x(F(y,\beta)&+G(y,\beta)\log(x)) +O(x^2),\\
	{W_3}^1= -\mathrm{i}\beta\pi+\mathrm{i}H(y,\beta)x +O(x^2)&,\,{W_3}^2= -\mathrm{i}\beta\pi-\mathrm{i}H(y,\beta)x +O(x^2),\\
	{W_4}^1= \mathrm{i}J(y,\beta)x +O(x^2)&,\,{W_4}^2= -\mathrm{i}J(y,\beta)x +O(x^2).\\
\end{split}
\end{equation}
where $F,G,H,I$ are regular when $x\rightarrow 0$. 
The only subtlety of the definition \eqref{eq:pinchedmatrix} is the logarithmic singularity in the contour integral over $\gamma_2$. It should not be surprised though, because when $x\rightarrow 0$, the branch cut disappear and the loop $\gamma_2$ get pinched. This sudden change implies that the derivative of the ${W_2}^{1(2)}$ with respect to $x$ at $x=0$ will not converge, more precisely, it diverges like $\log(x)$. This behavior of divergence is studied in detail in appendix \ref{sc:x=0}. One can also find similar examples in the logarithmic conformal field theory, see for example \cite{Flohr:2004ug,Gurarie:1993xq}.

 Fortunately this kind of singularity will not appear in the partition function. As it was suggested in \cite{multiloop}, the classical contribution will cancel the logarithmic singularity in the quantum part if one performs Poisson resummation of the terms in $S_{cl}$ which vanishes like $1/\log(x)$.
In appendix \ref{sc:x=0}, we show this calculation explicitly. We also find that the exact form of $F,G$ and $H$ are actually irrelevant for the leading singular term. In the end  $f(\beta)$ can be fixed as
\begin{equation}
f(\beta)=c_nR^6\vartheta_1^\prime(0)^{-1}|\eta(\mathrm{i}\beta)|^{-4},
\end{equation}
where we have absorbed other coefficients into $c_n$.

\begin{figure}[h]	
	\begin{subfigure}{0.5\textwidth}
		\includegraphics[width=0.9\linewidth, height=4cm]{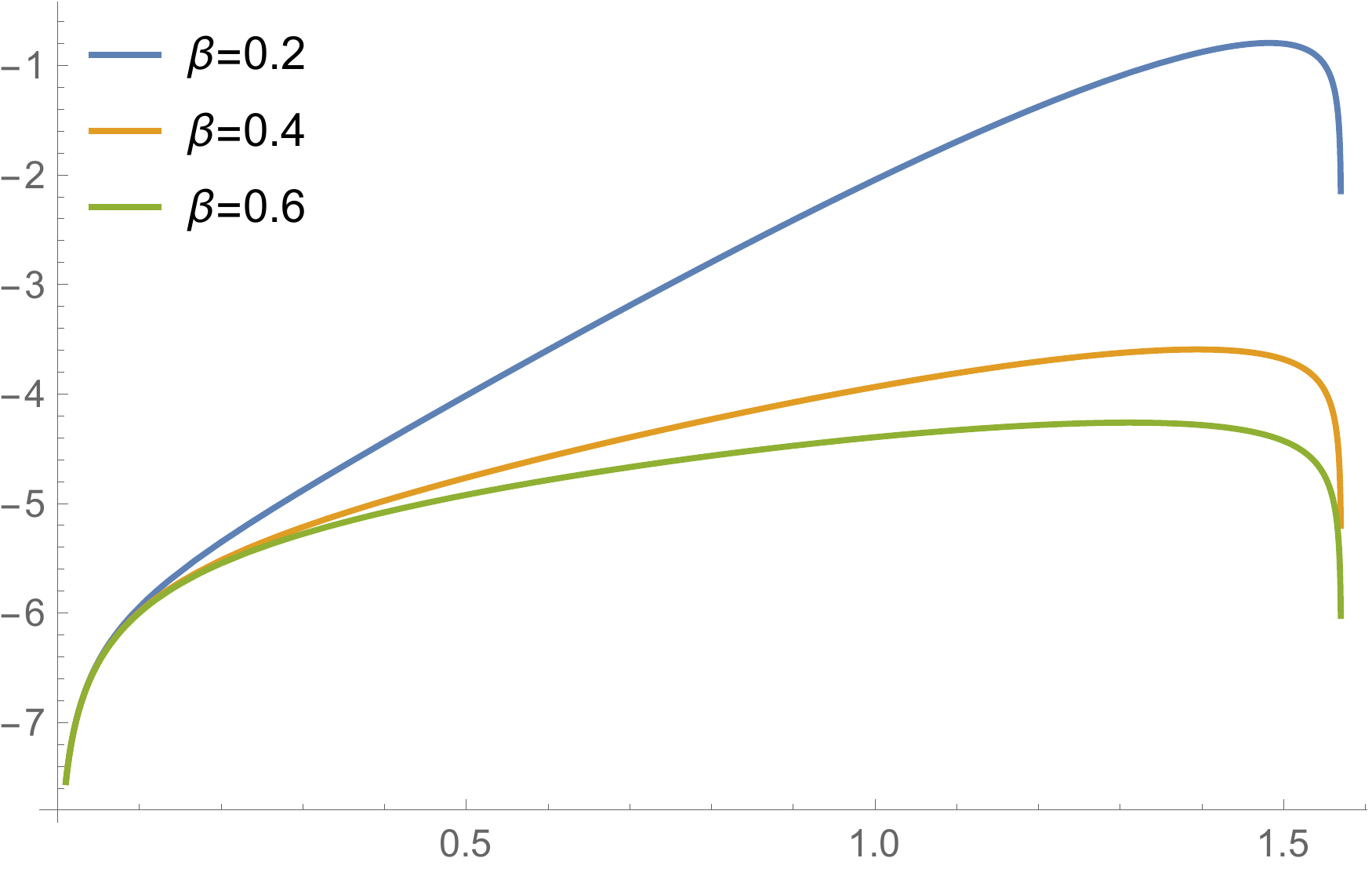} 
	\end{subfigure}
	\begin{subfigure}{0.5\textwidth}
		\includegraphics[width=0.9\linewidth, height=4cm]{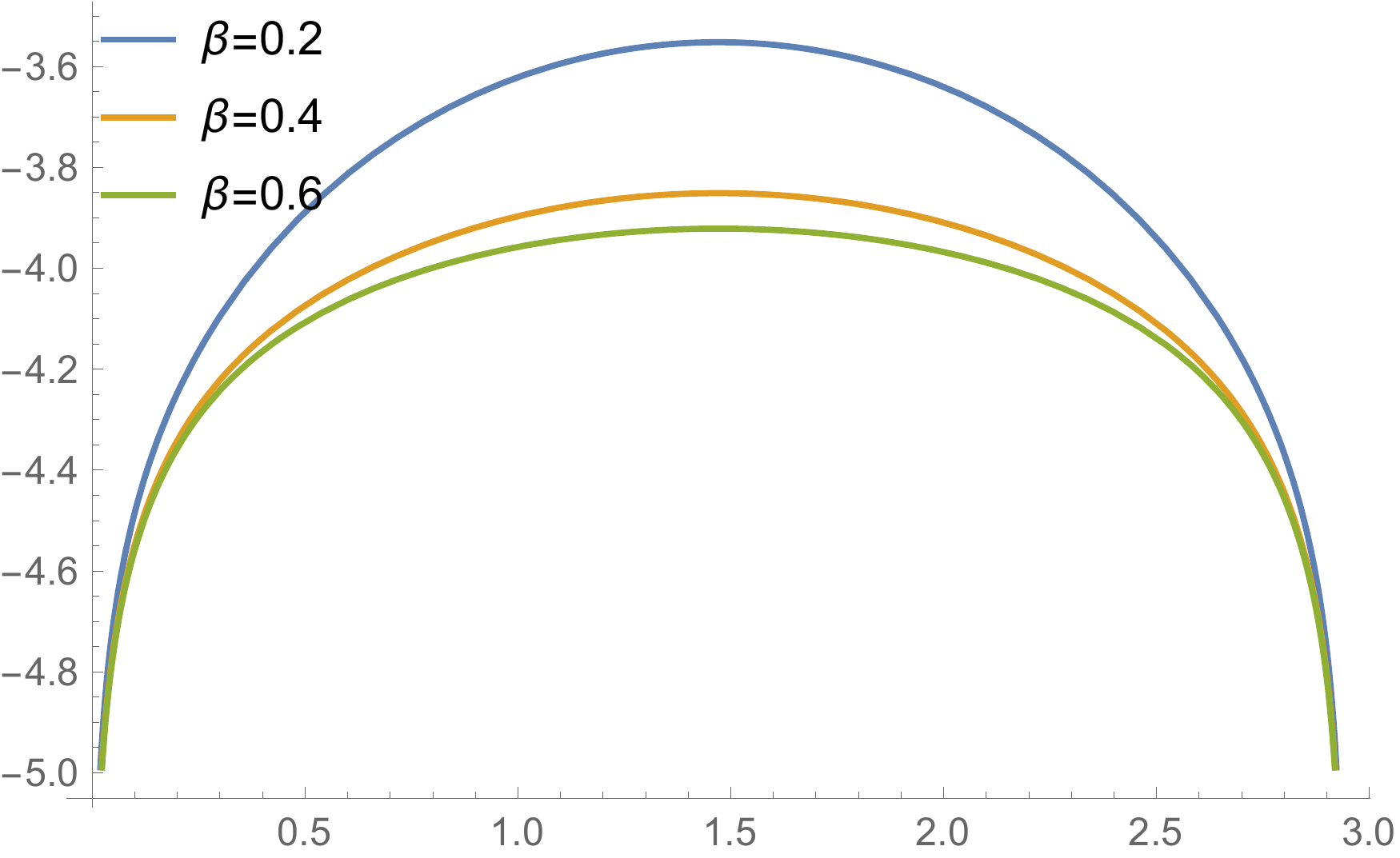}
	\end{subfigure}
	\caption{In the left we set $y=0.001$. In the right we set $x=0.1$.}
	\label{fig:entropy}
\end{figure}

Finally, the partition function $Z$ becomes
\begin{equation}\label{eq:results}
Z=c_n\frac{R^6}{\beta\vartheta_1^\prime(0)|\eta(\mathrm{i}\beta)|^8}\frac{1}{|\det{W}|}\frac{|\vartheta_1(x+y)||\vartheta_1(x)|}{\sqrt{|\vartheta_1(2x+y)||\vartheta_1(y)|}}\Theta(0|\frac{\mathrm{i}R^2}{4} \Gamma)^2 \Theta(0|\frac{\mathrm{i}R^2}{4} \Gamma^\prime)^2.
\end{equation}
This is the main result of the paper.
To obtain $Tr(\rho_A^N)$, $Z$ should be normalize with the original partition function
\begin{equation}\label{eq:totalZ}
Z_2\equiv Tr(\rho_A^N)=\frac{Z}{Z_b^2},
\end{equation}
Then the $N=2$ R\'enyi entropy is
\begin{equation}\label{eq:entropy}
\begin{split}
S&=-\log(c_n)-\log\left(\frac{1}{\beta\vartheta_1^\prime(0)}\frac{1}{|\det{W}|}\frac{|\vartheta_1(x+y)||\vartheta_1(x)|}{\sqrt{|\vartheta_1(2x+y)||\vartheta_1(y)|}}\right)\\
&\quad -\log\left(\Theta(0|\mathrm{i}R^2 \Gamma/4)^2 \Theta(0|\mathrm{i}R^2 \Gamma^\prime/4)^2\right)+4\log\left(\vartheta_3(0|\frac{\mathrm{i}2 \beta}{R^2})\right)\\
&\qquad+4\log\left(\vartheta_3(0|\frac{\mathrm{i}\beta R^2}{2})\right),
\end{split}
\end{equation}
This expression can be evaluated numerically for any $x\ne 0$ and $y\ne 0$. We plot \eqref{eq:entropy} as a function of $x$ and $y$ in figure \ref{fig:entropy}   

\subsubsection{T-duality}\label{sec:t}

It is important to check the T-duality of \eqref{eq:z}.  
Let's first draw out the $R$ dependent part of the partition function
\begin{equation}\label{eq:Rdependence}
\mathcal{F}(R)\equiv R^6\Theta(0| \mathrm{i} R^2\Gamma/4)^2 \Theta(0| \mathrm{i} R^2\Gamma^\prime/4)^2, 
\end{equation}
To condense the expression, we introduce the two functions 
\begin{equation}
D_{12}={W_1}^1{W_2}^2-{W_2}^1{W_1}^2, \,D_{34}={W_3}^1{W_4}^2-{W_4}^1{W_3}^2.
\end{equation}
The key observation is that $A,B,C$ and $A^\prime,B^\prime,C^\prime$ have a relation
\begin{equation}\label{eq:relation}
A=-2\frac{D_{34}}{D_{12}}C^\prime,\,B=\frac{D_{34}}{D_{12}}B^\prime,\,C=-\frac{D_{34}}{2D_{12}}A^\prime,
\end{equation}
and the following identity exists:
\begin{equation}
\frac{D_{34}}{2D_{12}}\frac{1}{(B^2-AC)}=1.
\end{equation}
These lead to an important relation
\begin{equation}\label{eq:niceequation}
4\Gamma^{-1}=\Gamma^{\prime}.
\end{equation}
It is also known that Riemann-Siegel theta function obeys the modular transformation \cite{NIST} 
\begin{equation}\label{eq:usefullformula}
\Theta(0|\Omega)=\Theta(0|-\Omega^{-1})\det{(\mathrm{-i}\Omega)}^{-1/2}.
\end{equation}
Therefore we have
\begin{equation}\label{eq:finalentropy}
\begin{split}
\Theta(0| \mathrm{i} R^2\Gamma^\prime/4)&= \Theta(0|\frac{ 4\mathrm{i}}{ R^2}{\Gamma^\prime}^{-1})\det(R^2\Gamma^\prime/2)^{-1/2}\\
&=\Theta(0|\frac{ \mathrm{i}}{ R^2}\Gamma)(\frac{4}{R^2})^{3/2}\det(\Gamma^\prime )^{-1/2}.
\end{split}
\end{equation}
Plugging into the equation \eqref{eq:Rdependence}
\begin{equation}
\label{eq:R2}
\mathcal{F}(R)\sim \Theta(0|\frac{\mathrm{i} R^2}{4}\Gamma)^2\Theta(0|\frac{ \mathrm{i}}{R^2}\Gamma)^2\det(\Gamma^\prime )^{-1},
\end{equation}
which is manifestly T-dual invariant $R^2\leftrightarrow \frac{4}{R^2}$.

\section{Low temperature expansion}\label{sc:lowt}
\begin{figure}[h]
	\centering
	\includegraphics[width=3in]{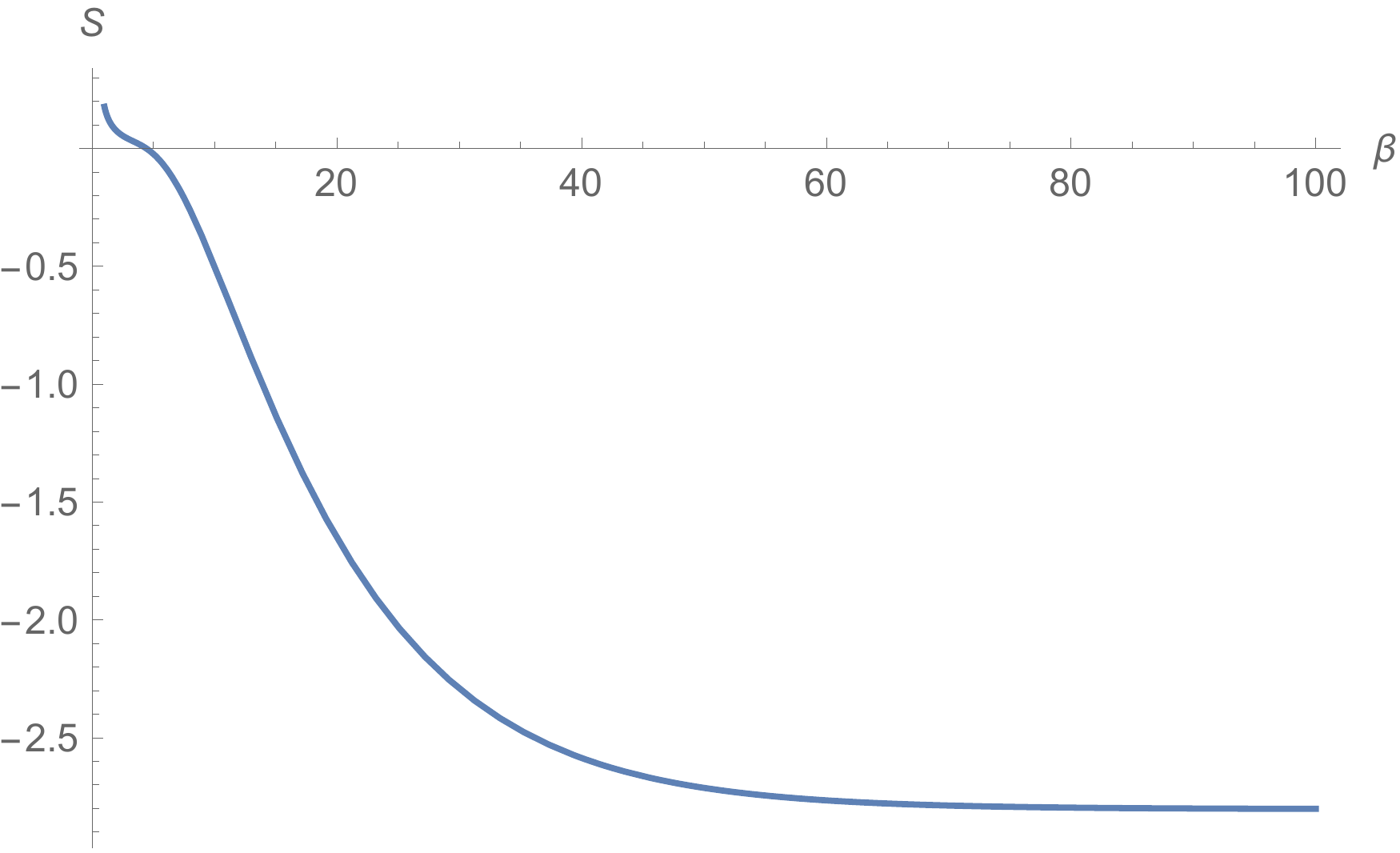}
	\caption{\label{fig:zerot} We have set $x=1$, $y=0.1$ and $R=1.5$}
\end{figure}

To see the temperature dependence of the R\'enyi entropy more clearly, we plot it in figure \ref{fig:zerot}.
It shows a zero temperature limit as we expected.
In order to find out what the low temperature limit is, we would like to expand the partition function \eqref{eq:totalZ} with respect to $q\equiv e^{-\beta \pi}$. The following expansion are useful:
\begin{equation}
\begin{split}
	\vartheta_1(z,q)&\sim 2 \sin(z)q^{1/4}+O(q^2)\\
	\vartheta^\prime_1(z,q)&\sim 2\cos(z)q^{1/4}+O(q^2)\\
	\vartheta_3(z,q)&\sim 1+ 2 \cos(z)q+O(q^2).
\end{split}
\end{equation}
In the limit $\beta\rightarrow\infty$, only the contour integral over the thermal cycle diverges as $-\mathrm{i}\beta\pi$. but all the other elements of the period matrix are finite, this can be seen by numerical evaluation or Taylor expansion of the integrand according to $q=e^{-\beta \pi}$. Hence we can introduce a general form of the leading term of the period matrix in the low temperature limit
\begin{equation}
W \sim
\left(
\begin{array}{cccc}
\pi+O(q^2)  & \pi+O(q^2)  & \pi+O(q^2)  & \pi+O(q^2)  \\
\text{$f_1$} \pi+O(q^2)  & \text{$f_2$} \pi+O(q^2)  & \text{$f_1$} \pi+O(q^2)  & \text{$f_2$} \pi+O(q^2)  \\
\pi  (i \text{$g_1$}-i \beta )+O(q^2) & \pi  (i \text{$g_2$}-i \beta ) +O(q^2)& \pi  (-i \text{$g_1$}+i \beta )+O(q^2) & \pi  (-i \text{$g_2$}+i \beta )+O(q^2) \\
i \text{$h_1$} \pi +O(q^2) & i \text{$h_2$} \pi +O(q^2) & -i \text{$h_1$} \pi +O(q^2) & -i \text{$h_2$} \pi  +O(q^2)\\
\end{array}
\right),
\end{equation}
where $f_{1(2)}$, $g_{1(2)}$ and $h_{1(2)}$ are some functions of $x$ and $y$.

For convenience, we transform the partition function \eqref{eq:totalZ} into its manifested T-dual invariant form
\begin{equation}
	\label{eq:tdualform}
Z_2=\frac{c_n}{\beta\vartheta_1^\prime(0)}\frac{1}{|\det{W}||\det{\Gamma^\prime}|}\frac{|\vartheta_1(x+y)||\vartheta_1(x)|}{\sqrt{|\vartheta_1(2x+y)||\vartheta_1(y)|}}\frac{ \Theta(0|\mathrm{i}\frac{R^2}{4}\Gamma)^2 \Theta(0|\frac{\mathrm{i}}{R^2} \Gamma)^2}{\vartheta_3(0|\frac{\mathrm{i}\beta R^2}{2})^4\vartheta_3(0|\frac{\mathrm{i}2\beta}{ R^2})^4}.
\end{equation}
After some algebra, we get
\begin{equation}
\lim_{\beta\rightarrow \infty}\beta|\det{W}||\det{\Gamma^\prime}|=32 \pi^4 (f_1-f_2)^2.
\end{equation}
Now we expand the Riemann-Siegel theta function in the large $\beta$ limit.
Given the form of $\Gamma$ matrix
\begin{equation}
\Gamma=
\begin{pmatrix}
\frac{f_2g_1-f_1g_2}{f_1-f_2}+2\beta & \frac{-f_2g_1+f_1g_2}{f_1-f_2} & \frac{-2g_1 +2 g_2+f_2h_1-f_1h_2}{2(f_1-f_2)}\\
\frac{-f_2g_1+f_1g_2}{f_1-f_2} & \frac{f_2g_1-f_1g_2}{f_1-f_2}+2\beta & \frac{2g_1-2 g_2-f_2h_1+f_1h_2}{2(f_1-f_2)}\\
\frac{-2g_1 +2 g_2+f_2h_1-f_1h_2}{2(f_1-f_2)} & \frac{2g_1-2 g_2-f_2h_1+f_1h_2}{2(f_1-f_2)} & 2\frac{2(h_1-h_2)}{f_1-f_2}
\end{pmatrix},
\end{equation} 
one can see that the leading contribution of the summation
\begin{equation}
\Theta(0|\frac{\mathrm{i}}{R^2} \Gamma)=\sum_{m_1,m_2,m_3}e^{-\frac{\pi}{R^2}m.\Gamma.m}
\end{equation}
comes from $m_1=m_2=0$, and the next leading term comes from $m_1^2+m_2^2=1$. 
For simplicity, we also assuming that $R^2>2$, then the expansion of Riemann-Siegel theta function is approximately
\begin{equation}
\begin{split}
\Theta(0|\frac{\mathrm{i}}{R^2}\Gamma)& \sim \sum_{m_3}e^{-\frac{2\pi }{R^2}\frac{h_2-h_1}{f_1-f_2}m_3^2}+
4e^{-\frac{2\pi}{R^2} \beta}\sum_{m_3}e^{-\frac{\pi }{R^2}\left(\frac{2(h_2-h_1)}{f_1-f_2}m_3^2+\lambda m_3+ \frac{(f_2g_1-f_1g_2)}{f1-f2}\right)}+O(e^{-\frac{4\pi}{R^2} \beta})\\
&=\vartheta_3(0|\frac{2\mathrm{i}}{R^2}\tilde{\beta})\left(1+4\frac{\vartheta_3(\frac{\lambda}{2R^2}|\frac{2\mathrm{i}}{R^2}\frac{1}{\tilde{\beta}})}{\vartheta_3(0|\frac{2\mathrm{i}}{R^2}\frac{1}{\tilde{\beta}})}e^{-\frac{\pi}{R^2}\frac{(f_2g_1-f_1g_2)}{f1-f2}}e^{-\frac{2\pi}{R^2} \beta} +O(e^{-\frac{4\pi}{R^2} \beta})\right)
\end{split}
\end{equation}
where we have defined
\begin{equation}
\lambda\equiv\frac{2g_1-2g_2-f_2h_1+f_1h_2}{f_1-f_2},\quad \frac{h_2-h_1}{f_1-f_2}\equiv  \frac{1}{\tilde{\beta}}.
\end{equation}
With these results, the partition function is approximately
\begin{equation}\label{eq:largebeta}
\begin{split}
Z_2&=\frac{c_n}{32 \pi^4 (f_1-f_2)^2}\frac{|\sin(x+y)||\sin(x)|}{\sqrt{|\sin(2x+y)||\sin(y)|}} \vartheta_3(0|\mathrm{i}\frac{2}{R^2}\frac{1}{\tilde{\beta}})^2\vartheta_3(0|\mathrm{i}\frac{R^2}{2}\frac{1}{\tilde{\beta}})^2\\
&\qquad \left(1+8\frac{\vartheta_3(\frac{\lambda}{2R^2}|\frac{2\mathrm{i}}{R^2}\frac{1}{\tilde{\beta}})}{\vartheta_3(0|\frac{2\mathrm{i}}{R^2}\frac{1}{\tilde{\beta}})}e^{-\frac{\pi}{R^2}\frac{(f_2g_1-f_1g_2)}{f_1-f_2}}e^{-\frac{2\pi}{R^2} \beta} +O(e^{-\frac{4\pi}{R^2} \beta})\right),
\end{split}
\end{equation}
where $f_1,f_2$ and $g_1,g_2$ are given as following
\begin{equation}\label{eq:smally}
	\begin{split}
		\pi f_1&=-2\int_{\frac{\pi+y}{2}}^{\frac{\pi-y}{2}} \ud z \frac{\sqrt{\cos(z+y/2)}\sqrt{\cos(z+y/2)}}{\sqrt{\cos(z-x-y/2)}\sqrt{\cos(z+x+y/2)}}\\
	\pi f_2&=-2\int_{\frac{\pi-y}{2}}^{\frac{\pi-y}{2}} \ud z \frac{\sqrt{\cos(z-x-y/2)}\sqrt{\cos(z+x+y/2)}}{\sqrt{\cos(z-y/2)}\sqrt{\cos(z+y/2)}}\\
	\pi g_1&=\int_{0}^{-i\beta \pi} \ud z \left[ \frac{\sqrt{\cos(z-y/2)}\sqrt{\cos(z+y/2)}}{\sqrt{\cos(z-x-y/2)}\sqrt{\cos(z+x+y/2)}}-1\right]\\
	\pi g_2&=\int_{0}^{-i\beta \pi} \ud z \left[ \frac{\sqrt{\cos(z-x-y/2)}\sqrt{\cos(z+x+y/2)}}{\sqrt{\cos(z-y/2)}\sqrt{\cos(z+y/2)}}-1\right],
	\end{split}
\end{equation}
and $h_1$ and $h_2$ can be defined similarly as $f_1$ and $f_2$ except for different contour.
It is interesting to notice that, the contour integrals along the canonical cycles of the torus drop out in the leading term, 
they however reappear in the sub-leading terms.

\subsection{Large system limit}
Now we focus on the leading term which only depends on $f_1-f_2$ and $h_2-h_1$:
\begin{equation}
\begin{split}
f_1-f_2&=\frac{1}{\pi}\oint_{\gamma_2}\ud z( w^1(z)-w^2(z))\\
h_1-h_2&=\frac{1}{\mathrm{i}\pi}\oint_{\gamma_4}\ud z( w^1(z)-w^2(z)).
\end{split}
\end{equation}
To condense the notations, let's define
\begin{equation}
\begin{split}
w(z)&=w^1(z+\pi/2)-w^2(z+\pi/2)\\
&=\frac{\vartheta_1(z-\frac{y}{2})\vartheta_1(z+\frac{y}{2})-\vartheta_1(z-\frac{y}{2}-x)\vartheta_1(z+\frac{y}{2}+x)}{\vartheta_1(z-\frac{y}{2})^{1/2}\vartheta_1(z+\frac{y}{2})^{1/2}\vartheta_1(z-\frac{y}{2}-x)^{1/2}\vartheta_1(z+\frac{y}{2}+x)^{1/2}}.
\end{split}
\end{equation}
In the large $\beta$ limit, $w(z)$ becomes
\begin{equation}
w(z)=\frac{\sin(z-\frac{y}{2})\sin(z+\frac{y}{2})-\sin(z-\frac{y}{2}-x)\sin(z+\frac{y}{2}+x)}{\sin(z-\frac{y}{2})^{1/2}\sin(z+\frac{y}{2})^{1/2}\sin(z-\frac{y}{2}-x)^{1/2}\sin(z+\frac{y}{2}+x)^{1/2}}+O(q^2)
\end{equation}

To compare the leading term of \eqref{eq:largebeta} with the earlier results in \cite{eeoftwo}, we further consider the infinite system limit: $x<<\pi$ and $y<<\pi$, i.e., the length of the subsystem and their separation are much smaller than that of the whole system. In this way the contour integral can be further simplified
\begin{equation}
\begin{split}
f_1-f_2&=\frac{1}{\pi}\oint_{\gamma_2}\ud z w(z)\\
&=\frac{1}{\pi}\oint_{\gamma_2}\ud z\frac{(x+y)x}{(z-\frac{y}{2})^{1/2}(z+\frac{y}{2})^{1/2}(z-\frac{y}{2}-x)^{1/2}(z+\frac{y}{2}+x)^{1/2}}.
\end{split}
\end{equation}
This integral is easily calculated giving
\begin{equation}
\begin{split}
&\oint_{\gamma_2}\ud z \frac{1}{(z-\frac{y}{2})^{1/2}(z+\frac{y}{2})^{1/2}(z-\frac{y}{2}-x)^{1/2}(z+\frac{y}{2}+x)^{1/2}}=\frac{2\pi\mathrm{i}F_{1/2}(1-r)}{(x+y)}\\
&\oint_{\gamma_4}\ud z \frac{1}{(z-\frac{y}{2})^{1/2}(z+\frac{y}{2})^{1/2}(z-\frac{y}{2}-x)^{1/2}(z+\frac{y}{2}+x)^{1/2}}=\frac{2\pi\mathrm{i}F_{1/2}(r)}{(x+y)},
\end{split}
\end{equation}
where we have defined
\begin{equation}
r=\frac{(z_1-z_2)(z_3-z_4)}{(z_1-z_3)(z_2-z_4)},\, F_{1/2}(r)={}_2F_1(1/2,1/2;1;r).
\end{equation}
Thus we have
\begin{equation}
\begin{split}
f_1-f_2&=2\mathrm{i}F_{1/2}(1-r)x\\
h_2-h_1&=2\mathrm{i}F_{1/2}(r)x,
\end{split}
\end{equation}
and
\begin{equation}
\begin{split}
\tilde{\beta}^{-1}=\frac{F_{1/2}(r)}{F_{1/2}(1-r)}
\end{split}
\end{equation}
At last, the leading term of \eqref{eq:largebeta} can be written as
\begin{equation}\label{eq:lowtlimit}
\begin{split}
Z_2&=c_n\left[\frac{(x+y)x}{x^2\sqrt{(2x+y)y}}\right]\frac{1}{F_{1/2}(1-r)^2} \vartheta_3(0|\mathrm{i}\frac{2}{R^2}\frac{1}{\tilde{\beta}})^2\vartheta_3(0|\mathrm{i}\frac{R^2}{2}\frac{1}{\tilde{\beta}})^2\\
&=c_n\left[\frac{x+y}{x\sqrt{(2x+y)y}}\right]\frac{1}{F_{1/2}(1-r)^2\tilde{\beta}^{-2}} \vartheta_3(0|\mathrm{i}\frac{2}{R^2}\tilde{\beta})^2\vartheta_3(0|\mathrm{i}\frac{R^2}{2}\tilde{\beta})^2\\
&=c_n\left[\frac{x+y}{x\sqrt{(2x+y)y}}\right]\frac{1}{F_{1/2}(r)^2} \vartheta_3(0|\mathrm{i}\frac{2}{R^2}\tilde{\beta})^2\vartheta_3(0|\mathrm{i}\frac{R^2}{2}\tilde{\beta})^2\\
&=c_n\left[\frac{x+y}{x\sqrt{(2x+y)y}}\right]\left[\frac{\vartheta_3(0|\mathrm{i}\frac{2}{R^2}\tilde{\beta})\vartheta_3(0|\mathrm{i}\frac{R^2}{2}\tilde{\beta})}{\vartheta_3^2(\tilde{\beta})} \right]^2
\end{split}
\end{equation}
where we have used the equality \cite{eeoftwo}
\begin{equation}
F_{1/2}(r)=\vartheta_3^2(\tilde{\beta}).
\end{equation}
One can see that the result \eqref{eq:lowtlimit} is agreed with \cite{eeoftwo} for the $N=2$ case.

\subsection{Universal thermal corrections in the limit of small seperation}

As one can see that the thermal corrections in the expansion \eqref{eq:largebeta} is very complicate.
In order to compare it with the results for a single interval case \cite{Chen:2015cna},
we consider a special case that the length of the two intervals are much 
bigger than the separation, i.e., $x\gg y$.
For further convenience, we also change the variable $\omega=e^{\mathrm{i}2z}$, so the integral \eqref{eq:smally} can be written as
\begin{equation}
\begin{split}
\pi f_1&=\mathrm{i}\int_{-e^{-\mathrm{i}y}}^{-e^{\mathrm{i}y}} \ud \omega \frac{\sqrt{1+2\omega\cos(y)+\omega^2}}{\omega\sqrt{1+2\omega\cos(2x+y)+\omega^2}}\\
\pi f_2&=\mathrm{i}\int_{-e^{-\mathrm{i}y}}^{-e^{\mathrm{i}y}} \ud \omega \frac{\sqrt{1+2\omega\cos(2x+y)+\omega^2}}{\omega\sqrt{1+2\omega\cos(y)+\omega^2}}\\
\pi g_1&=\frac{1}{2\mathrm{i}}\int_{1}^{0} \ud \omega\frac{1}{\omega}\left[ \frac{\sqrt{1+2\omega\cos(y)+\omega^2}}{\sqrt{1+2\omega\cos(2x+y)+\omega^2}}-1\right]\\
\pi g_2&=\frac{1}{2\mathrm{i}}\int_{1}^{0} \ud \omega\frac{1}{\omega}\left[ \frac{\sqrt{1+2\omega\cos(2x+y)+\omega^2}}{\sqrt{1+2\omega\cos(y)+\omega^2}}-1\right].
\end{split}
\end{equation}
Since there are no pinching divergences in these integrals, we can safely Taylor expand the integrand with respect to $y$, and after that we do the integration, we find 
\begin{equation}
\begin{split}
\pi f_1\sim 0+O(y^2)&,\, \pi f_2\sim 2y+O(y^2)\\
\pi g_1=\log\left(\frac{1+\cos(2x)}{2}\right)-\tan(x)y+O(y^2)&,\, \pi g_2=xy+O(y^2).
\end{split}
\end{equation}
Then we get
\begin{equation}
\begin{split}
\frac{f_2g_1-f_1g_2}{f_1-f_2}\sim-\frac{1}{\pi}\log\left(\frac{1+\cos(2x)}{2}\right),\,
\end{split}
\end{equation}
On the other hand, $1/\tilde{\beta}$ diverges as $y\rightarrow 0$. 
Using the expansion
\begin{equation}
\vartheta_3(z,q)\sim 1+4\cos(2z)q+O(q^2),
\end{equation}
 the first oder of the thermal correction is approximately
\begin{equation}
\begin{split}
8\frac{\vartheta_3(\frac{\lambda}{2R^2}|\frac{2\mathrm{i}}{R^2}\frac{1}{\tilde{\beta}})}{\vartheta_3(0|\frac{2\mathrm{i}}{R^2}\frac{1}{\tilde{\beta}})}e^{-\frac{\pi}{R^2}\frac{(f_2g_1-f_1g_2)}{f_1-f_2}}e^{-\frac{2\pi}{R^2} \beta}&\sim 8 e^{\frac{1}{R^2}\log(\frac{1+\cos(2x)}{2})}e^{-\frac{2\pi}{R^2}\beta}\\
&\sim 8 (\frac{1+\cos(2x)}{2})^{\frac{1}{R^2}}e^{-\frac{2\pi}{R^2}\beta}.
\end{split}
\end{equation}
By using the identity
\begin{equation}
\frac{1+\cos(2x)}{2}=\cos^2(x)=\left(\frac{\sin(2x)}{2\sin(x)}\right)^2,
\end{equation}
the first order thermal correction of the partition function is just
\begin{equation}
8(\frac{\sin(2x)}{2\sin(x)})^{\frac{2}{R^2}}e^{-\frac{2\pi}{R^2}\beta},
\end{equation}
which is the same results for a single interval in case of $N=2$ \cite{Chen:2015cna,Cardy:2014jwa}, where $2x$ is just the 
total length of the intervals.

\section{Conclusion}\label{sc:conclusion}

In this paper we calculate the $N=2$ R\'enyi entanglement entropies of two intervals on a circle at finite temperature \eqref{eq:entropy}. We also obtain the low temperature expansion up to the second order with respect to $e^{-\frac{2\pi}{R^2}\beta}$. A non-trivial check is made by taking the large system limit and the leading term is agreed with the R\'enyi entanglement entropy of two intervals in an infinite system at zero temperature \cite{eeoftwo}.
Furthermore, when we take the small separation limit $y\ll x$, the low temperature expansion also gives the correct universal thermal corrections for a single interval.

As we have seen, the quantum part of the partition function is essentially the four point function of twist fields. The very interesting thing about the twist fields is that they create branch cuts on the Riemann surface. This is why we encounter some subtle logarithmic term in the quantum part as we colliding the twist/antitwist pairs of operators. However this logarithmic behavior doesn't show in the two point function \cite{1311.1218,Chen:2015cna}, this implies that when we calculating the four point function, there are actually two independent conformal block depending on the choice of internal channel in the OPE, i.e., the different screening contours which will get pinched or not. Fortunately, for compact boson this logarithmic singularity is canceled by the classical contribution in the end. Nevertheless this logarithmic behavior is interesting for their own sake, which deserves further study.


\section{Acknowledgement}
We would like to thank Wei Fu for useful discussion and our colleagues at the University of Electronic Science and Technology of China for kind support. This work is supported by various generous grants, whose official names we hope to find out soon.

\appendix

\section{Inner products of cut abelian differentials}
\label{sc:innerproduct}
\begin{figure}[h]
   \centering
  \includegraphics[width=4in]{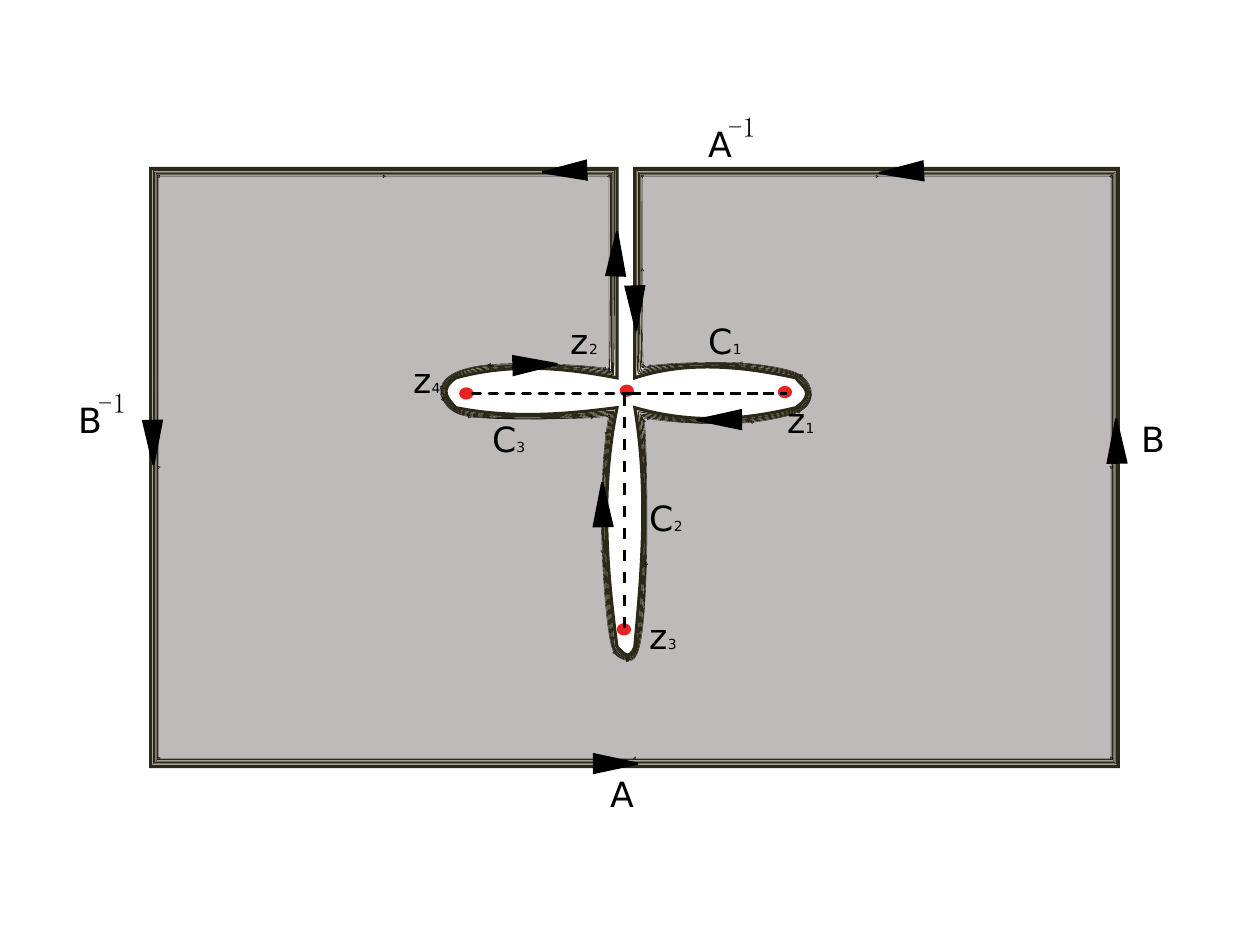}
  \caption{\label{fig:cut1} We have chosen a convenient contour which encircle all the branch points and cuts. Also the path $C_a$ are closed related to the basis of loops $\gamma_a$.}
\end{figure}

The inner product are defined by
\begin{equation}
  \label{eq:innerproduct}
  (w^{i}, w^{j}) \equiv \mathrm{i}\int_R w^{i} \wedge \bar w^{j}, \quad w^i=w^i(z) dz.
\end{equation}
Following the same strategy in \cite{tatalecture}, where it was used to prove Riemann bilinear relation, one can show that the inner product can be calculated by doing contour integral along the edges of the shadow region as depicted in figure \ref{fig:cut1}.
Since $w^l$ is a holomorphic one form on the region $\Pi$, one can always find a holomorphic function $f^l$ such that $\omega^l=df^l$. By Stoke's theorem, the inner product can be written as a contour integral on the boundaries
\begin{equation}
  \label{eq:contour}
  \begin{split}
\frac{1}{\mathrm{i}}(w^i,w^j)&=\oint_{\partial \Pi} f^i\bar w^j=  \int_{A} w^i\int_{B} \bar w^j-\int_{B}w^{i} \int_{A} \bar w^{j}\\
	&+\int_{C_1}w^i\int_{C_2}w^j+\int_{C_2}w^i\int_{C_3}w^j+\int_{C_1}w^i\int_{C_3}w^j\\
	&+\frac{1}{1-e^{-2\pi k/N}}\int_{C_1}w^i\int_{C_1}w^j+\frac{1}{1-e^{-2\pi k/N}}\int_{C_2}w^i\int_{C_2}w^j\\
	&+\frac{1}{1-e^{2\pi k/N}}\int_{C_3}w^i\int_{C_3}w^j
  \end{split}
\end{equation}
Given the relation $\int_{C_3}+\int_{C_2}+\int_{C_1}=0$ and
\begin{equation}
\int_{C_1}=-\oint_{\gamma_4}, \qquad \int_{C_2}=-\oint_{\gamma_2},
\end{equation}
 the inner product for $k=1, N=2$ can be presented by the elements of cut period matrix
\begin{equation}
\label{eq:innerproduct1}
(w^{i}, w^{j})=-\mathrm{i}(W_1^{i}\bar W_3^{j}-W_3^{i}\bar W_1^{j})
	+\frac{\mathrm{i}}{2}\left(W_4^i \bar W_2^j-W_2^i \bar W_4^j \right).
\end{equation}
It is easy to verify that the inner product is hermitian.

\section{Definition of contour integrals}
\label{sc:proof}
The convention of theta functions we used is the same as \cite{NIST}:
\begin{equation}
\label{eq:convention}
\begin{split}
	\vartheta_1(z|\tau)&=\vartheta_1(z,q)\\
		&=2\sum_{n=0}^{\infty}(-1)^n q^{(n+1/2)^2}\sin((2n+1)z).
\end{split}
\end{equation}
where $q\equiv e^{i\pi \tau}$.
The theta function are quasi-periodic
\begin{equation}
\vartheta_1(z+(m+n\tau)\pi|\tau)=(-1)^{(m+n)}q^{-n^2}e^{-2\mathrm{i}nz}\vartheta_1(z|\tau).
\end{equation}

\begin{figure}[h]
	\centering
	\includegraphics[width=4in]{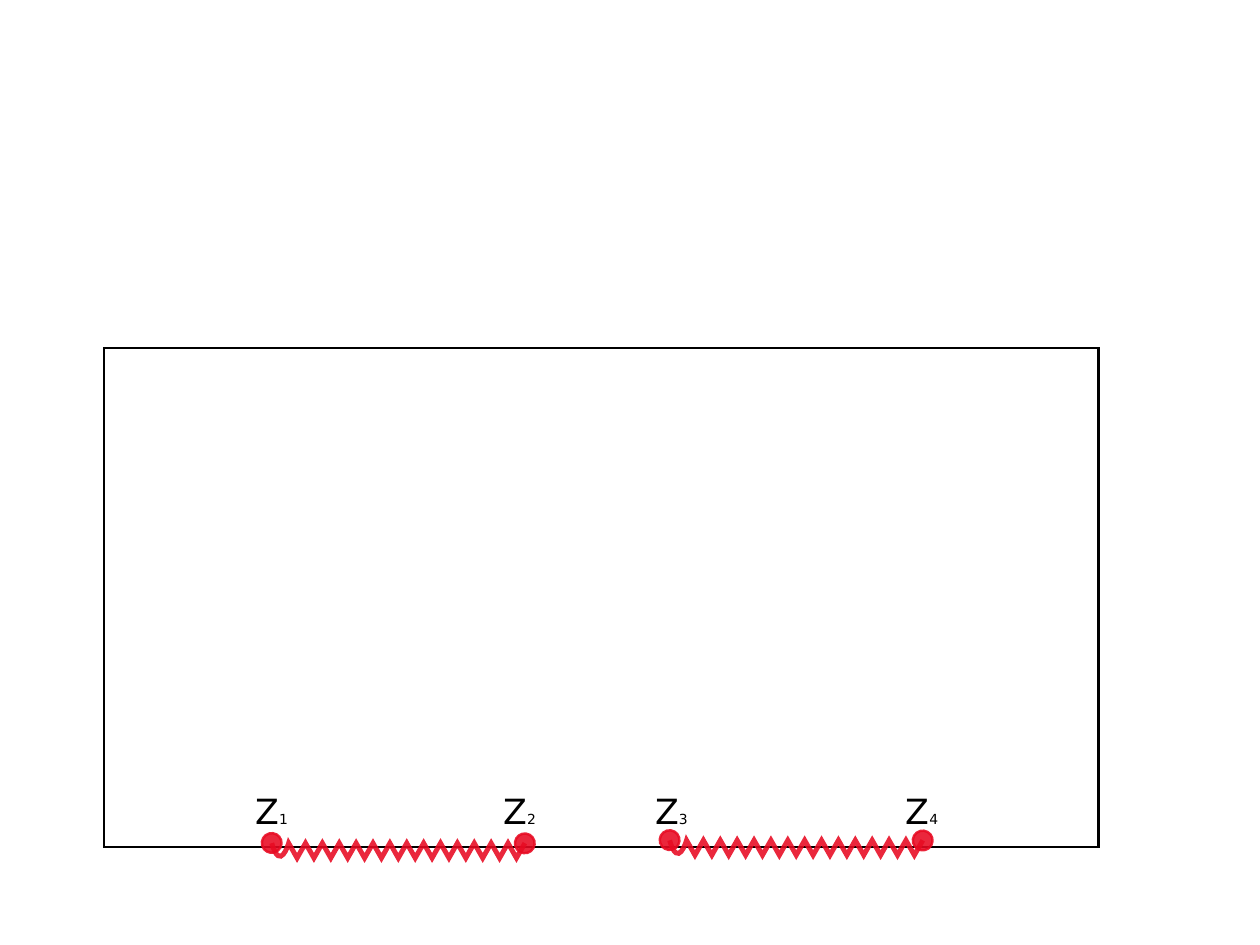}
	\caption{The two branch cuts\label{fig:cut}}
\end{figure}
For simplicity, we put the branch points on the real axis, see figure \ref{fig:cut}. The contour integral is defined as follows.
Since the closed loop circling the four twist insertion is trivial, so that the integral along $(z_1,z_2)$ and $(z_3,z_4)$ will cancel each other.
Then $W_1^1$ can be written as 
\begin{equation}
{W_1}^{1(2)}(x,y)=(\int_{0}^{z_1}\ud z+\int_{z_2}^{z_3}\ud z+\int_{z_4}^{\pi}\ud z)w^{1(2)}(z).
\end{equation}
Since the theta function $\vartheta_1(z)$ is an odd function, and if $\tau$ is pure imaginary, $\vartheta_(z,q)$ is always real on the real line, which indicate that the integral $W_1^1$ and $W_1^2$ are real.

To do the contour integral around $\gamma_2$, we chose the branch to be $(-\pi,\pi)$. Then 
\begin{equation}
W_2^{1(2)} =\oint_{\gamma_2} \ud z \omega^1(z)=(e^{2 \pi \mathrm{i}\frac{1}{2}}-1)\int_{z_2}^{z_3}\ud z w^{1(2)}(z).
\end{equation}

While long the B-cycle $\gamma_3$, if we let $\tau=\mathrm{i} \beta$ to be pure imaginary, the contour integral can be written as
\begin{equation}
\begin{split}
{W_3}^{1(2)} =\oint_{\gamma_3} \ud z w^{1(2)}(z)=\int_{0}^{-\mathrm{i} \beta \pi }w^{1(2)}\ud z 
\end{split}
\end{equation}
The contour integral $W_4^{1(2)}$ are given by similarly
 \begin{equation}
 \begin{split}
 {W_4}^1 &=e^{-\frac{\mathrm{i} \pi}{2}}2 \mathrm{i} \sin(\frac{3}{2}\pi)(-1)^{-1/2}\int_{z_1}^{z_2}\ud z \vartheta_1(z_1-z)^{-1/2}\vartheta_1(z-z_2)^{1/2} \vartheta_1(z-z_3)^{1/2}\vartheta_1(z-z_4)^{-1/2}\\
 {W_4}^2 &=e^{-\frac{\mathrm{i} \pi}{2}}2 \mathrm{i} \sin(\frac{3}{2}\pi)(-1)^{1/2}\int_{z_1}^{z_2}\ud z \vartheta_1(z_1-z)^{1/2}\vartheta_1(z-z_2)^{-1/2} \vartheta_1(z-z_3)^{-1/2}\vartheta_1(z-z_4)^{1/2}
 \end{split}
 \end{equation}
The other element of period matrix can be obtained by doing conjugation.

\section{$x\rightarrow 0$ limit of contour integral}\label{sc:x=0}

In the following we will study the behavior of the contour integral as $x\rightarrow 0$.
The most important thing is 
to convince ourself that there is a logarithmic divergence of ${W_2}^{1(2)}$ when the loop $\gamma_2$ get pinched.
Noted that, in the limit $x\rightarrow0$, it is not helpful to expanded the integrand of ${W_2}^{1(2)}$
with respect to $x$, because we expect that the derivative of the integral is 
not regular at $x=0$. Therefore we should change the strategy. Since the singularity only depends on how close the two branch points get when pinching the contour, i.e., the singularity should not depends on $y$, thus it is enough to study the singularity by considering a much simpler case that $y$ is very small. By using the approximation that
\begin{equation}
\vartheta_1(z)\sim z
\end{equation}
when $z$ is small, the integral ${W_2}^{1}$ can be simplified:
\begin{equation}\label{eq:w21sim}
\begin{split}
{W_2}^{1}&\sim(-2)\int_{-y/2}^{+y/2}\ud z\frac{\sqrt{z-\frac{y}{2}-x}\sqrt{z+\frac{y}{2}+x
	}}{\sqrt{z-\frac{y}{2}}\sqrt{z+\frac{y}{2}}}\\
	&=-y\int_{-1}^{1}\ud u\frac{\sqrt{u-\frac{2x}{y}-1}\sqrt{u+\frac{2x}{y}+1
		}}{\sqrt{u-1}\sqrt{u+1}}\\
		&=-2y\int_{0}^{1}\ud \sin(\theta) \frac{\sqrt{\sin(\theta)-\frac{2x}{y}-1}\sqrt{\sin(\theta)+\frac{2x}{y}+1
			}}{\sqrt{\sin(\theta)-1}\sqrt{\sin(\theta)+1}}\\
			&=-2(y+2x)\int_{0}^{\pi/2}\ud \theta \sqrt{1-\frac{1}{(1+2x/y)^2}\sin^2\theta}\\
			&=-2(y+2x)\mathrm{E}(\frac{1}{(1+2x/y)^2})
\end{split}
\end{equation}
where $\mathrm{E(m)}$ is the second kind of elliptic integral defined by
\begin{equation}
\mathrm{E}(m)\equiv\int_0^{\pi/2}\ud \theta\sqrt{1-m^2\sin^2\theta}.
\end{equation}
We can expand \eqref{eq:w21sim} with respect to $x$
\begin{equation}\label{eq:smallxexpansion}
{W_2}^{1}\sim-2y+2\left(-1-2\log(2)-\log(y)+\log(x)\right)x+O(x^2),
\end{equation}
which gives us correct leading term and also confirms the existence of $\log(x)$ in the next leading term.
For bigger $y$, the integral will not be so simple as \eqref{eq:smallxexpansion}. Nevertheless, giving the fact that the pinching process only depends on $x$,  it is reasonable to assuming that the general form of expansion at order $O(x)$ should be like
\begin{equation}
2(F(y,\beta)+G(y,\beta)\log(x))x.
\end{equation}
where $F(y,\beta)$ and $G(y,\beta)$ can be easily evaluated numerically but we can't find the analytical form. Fortunately, $F$ and $G$ will not appear in the partition function, as we will show in the following.

In \cite{multiloop}, it was suggested that the classical contribution will cancel the logarithmic singularity in the quantum part. Given the general form of \eqref{eq:pinchedmatrix}, let's compute the classical contribution
\begin{equation}
Z_{cl}=\Theta(0|\frac{\mathrm{i}R^2}{4} \Gamma)^2 \Theta(0|\frac{\mathrm{i}R^2}{4} \Gamma^\prime)^2.
\end{equation}
We first note that the matrix $\Gamma$ looks like
\begin{equation}
\Gamma=\left(
\begin{array}{ccc}
2 \beta -\frac{H y}{\pi  F+G \pi  \log (x)} & \frac{H y}{\pi  F+G \pi  \log (x)} & -\frac{2 (\pi  H+J y)}{4 \pi  F+4 G \pi  \log (x)} \\
\frac{H y}{\pi  F+G \pi  \log (x)} & 2 \beta -\frac{H y}{\pi  F+G \pi  \log (x)} & \frac{2 (\pi  H+J y)}{4 \pi  F+4 G \pi  \log (x)} \\
-\frac{2 (\pi  H+J y)}{4 \pi  F+4 G \pi  \log (x)} & \frac{2 (\pi  H+J y)}{4 \pi  F+4 G \pi  \log (x)} & -\frac{J}{F+G \log (x)} \\
\end{array}
\right).
\end{equation}
One can use the formula \eqref{eq:usefullformula} to do the trick of Poisson resummation
\begin{equation}
\Theta(0|\frac{\mathrm{i}R^2}{4} \Gamma)=\Theta(0|\frac{\mathrm{i}4}{R^2} \Gamma^{-1})\det(\frac{R^2}{4}\Gamma)^{-1/2}.
\end{equation}
By using \eqref{eq:niceequation}, we find
\begin{equation}
Z_{cl}=\Theta(0|\frac{\mathrm{i}}{R^2} \Gamma^\prime)^2 \Theta(0|\frac{\mathrm{i}R^2}{4} \Gamma^\prime)^2\det(\frac{R^2}{4}\Gamma)^{-1}.
\end{equation}
Then the only term may contain logarithmic singularity is
\begin{equation}\label{eq:pinchqu}
\begin{split}
\det(W)^{-1}\det(\Gamma)^{-1}&\sim-\frac{F+G \log (x)}{32 \beta ^2 J x^2 \left(J \left(4 \pi ^2 \beta  F+J y^2\right)+4 \pi ^2 \beta  G J \log (x)+\pi ^2 H^2-2 \pi  H J y\right)}\\
 &\sim\frac{-1}{128J^2x^2\beta^3\pi^2}
\end{split}
\end{equation}
One can see that the $F,G$ and $\log(x)$ don't appear in the end.

To find the residue of the leading singular term
\begin{equation}
\lim_{x\rightarrow 0}Z\sim \frac{a}{x},
\end{equation}
we also need to calculate the leading contribution of
\begin{equation}
\Theta(0|\frac{\mathrm{i}}{R^2} \Gamma^\prime)^2 \Theta(0|\frac{\mathrm{i}R^2}{4} \Gamma^\prime)^2.
\end{equation}
Given the form of $\Gamma^\prime$ up to the order $O(x)$
\begin{equation}
\Gamma^\prime=\left(
\begin{array}{ccc}
\frac{2}{\beta } & 0 & -\frac{\pi  H+J y}{J \pi  \beta } \\
0 & \frac{2}{\beta } & \frac{\pi  H+J y}{J \pi  \beta } \\
-\frac{\pi  H+J y}{J \pi  \beta } & \frac{\pi  H+J y}{J \pi  \beta } & \frac{4 (-H y+F \pi  \beta +G \pi  \beta  \log (x))}{J \pi  \beta } \\
\end{array}
\right),
\end{equation}
the leading contribution is
\begin{equation}
Z_{cl}\sim \left[\sum_{m_1,m_2} \exp\left(\frac{-\pi}{R^2}\frac{2}{\beta}(m_1^2+m_2^2)\right)\right]^2\left[\sum_{n_1,n_2} \exp\left(\frac{-\pi R^2}{4}\frac{2}{\beta}(n_1^2+n_2^2)\right)\right]^2.
\end{equation}
After doing Poisson resummation of $m_1$ and $m_2$, $Z_{cl}$ can be written 
as
\begin{equation}\label{eq:pinchcl}
	\begin{split}
	Z_{cl}&\sim \frac{16\beta^2 }{R^2}\left[\sum_{m_1,m_2} \exp\left(\frac{-R^2 \beta \pi}{2}(m_1^2+m_2^2)\right)\right]^2\left[\sum_{n_1,n_2} \exp\left(\frac{-\pi R^2}{2}\frac{1}{\beta}(n_1^2+n_2^2)\right)\right]^2\\
	&=\frac{16\beta^2 }{R^2}\left[\sum_{m_1,n_1} \exp\left(\frac{-R^2 \pi}{2}(\beta m_1^2+\frac{n_1^2}{\beta})\right)\right]^2\left[\sum_{m_2,n_2} \exp\left(\frac{-R^2 \pi}{2}(\beta m_2^2+\frac{n_2^2}{\beta})\right)\right]^2.
	\end{split}
\end{equation}
Hence the most singular term of the total partition function when $x\rightarrow 0$ is
\begin{equation}\label{eq:smallx}
\begin{split}
Z&\sim
\frac{f(\beta)}{\beta|\eta(\mathrm{i}\beta)|^4}\frac{1}{8J^2x^2R^2\beta}\frac{|\vartheta_1(x+y)||\vartheta_1(x)|}{\sqrt{|\vartheta_1(2x+y)||\vartheta_1(y)|}}\left[\sum_{m_1,n_1} \exp\left(\frac{-R^2 \pi}{2}(\beta m_1^2+\frac{n_1^2}{\beta})\right)\right]^4\\
&=\frac{1}{x}\frac{f(\beta)}{\beta^2|\eta(\mathrm{i}\beta)|^4}\frac{\vartheta_1^\prime(0)}{8J^2R^2}
\left[\sum_{m_1,n_1} \exp\left(\frac{-R^2 \pi}{2}(\beta m_1^2+\frac{n_1^2}{\beta})\right)\right]^4,
\end{split}	
\end{equation} 
where $J$ can be easily determined by expand the integrand with respect to $x$
\begin{equation}
\begin{split}
	w^1(z)&=1+\frac{1}{2}\left(\frac{\vartheta_1^\prime(z-\frac{\pi}{2}-\frac{y}{2})}{\vartheta_1(z-\frac{\pi}{2}-\frac{y}{2})}-\frac{\vartheta_1^\prime(z-\frac{\pi}{2}+\frac{y}{2})}{\vartheta_1(z-\frac{\pi}{2}+\frac{y}{2})}\right)x+O(x^2)\\
	w^2(z)&=1-\frac{1}{2}\left(\frac{\vartheta_1^\prime(z-\frac{\pi}{2}-\frac{y}{2})}{\vartheta_1(z-\frac{\pi}{2}-\frac{y}{2})}-\frac{\vartheta_1^\prime(z-\frac{\pi}{2}+\frac{y}{2})}{\vartheta_1(z-\frac{\pi}{2}+\frac{y}{2})}\right)x+O(x^2).
\end{split}
\end{equation}
then $J$ can be found by picking up the residue
\begin{equation}
\begin{split}
\oint_{\gamma_4}\frac{1}{2}\left(\frac{\vartheta_1^\prime(z-\frac{\pi}{2}-\frac{y}{2})}{\vartheta_1(z-\frac{\pi}{2}-\frac{y}{2})}-\frac{\vartheta_1^\prime(z-\frac{\pi}{2}+\frac{y}{2})}{\vartheta_1(z-\frac{\pi}{2}+\frac{y}{2})}\right)x=\mathrm{i}Jx=\mathrm{i}\pi x.
\end{split}
\end{equation}

Comparing \eqref{eq:smallx} with $Z_b^2/x$:
\begin{equation}
Z_b^2/x=\frac{R^4}{4x}\frac{1}{\mathrm{Im}(\tau)^2|\eta(\tau)|^8}\left[\sum_{m,m^\prime}\exp\left(-\frac{\pi R^2|m\tau-m^\prime|^2}{2 \mathrm{Im}(\tau)}\right)\right]^4,
\end{equation}
$f(\beta)$ should be fixed by
\begin{equation}
f(\beta)=\frac{2\pi^2R^6}{\vartheta_1^\prime(0)|\eta(\mathrm{i}\beta)|^4}.
\end{equation}

\section{Calculate classical contribution by using orbifold}\label{sc:oldmethod}

Generally, one consider the twist/antitwist pairs 
of insertions of twist fields for a given $k\in \{0, 1, \cdots, N-1\}$.
Following \cite{eeoftwo, cft-of-orbifold}, the shifts in the global monodromy condition lies in the subset of 
a complicated lattice describe as
\begin{equation}
\label{eq:lattice}
\Lambda_{k/N}=\left\lbrace q=\pi R\sum_{j=0}^{N-1} e^{2 \pi \mathrm{i} jk/N} (m+\mathrm{i}n)\right\rbrace, \, v_a\in(1-e^{2 \pi \mathrm{i} k/N}) \xi_a.
\end{equation}
where $\xi_a \in \Lambda_{k/N}$.
The shift vectors $v_2,v_4$ is given by
\begin{equation}
\label{eq:shift}
v_2=(1-e^{2 \pi \mathrm{i} k/N}) \xi_2,\, v_4=(1-e^{2 \pi \mathrm{i} k/N}) \xi_4.
\end{equation}
For shift vectors $v_1$ and $v_3$, which are corresponding to the two cycles of the torus, should be parameterized by 
\begin{equation}
v_1=\xi_1,\, v_3=\xi_3.
\end{equation}

In the case of $N=2$ and $k=1$,
\begin{equation}
\xi_a=\pi R\left[(m_0^a-m_1^a)+\mathrm{i} (n_0^a-n_1^a)\right],\quad m_0^a,n_0^a,m_1^a,n_1^a \in \mathbb{Z}.
\end{equation}
Then the first half of the classical action is given by
\begin{equation}\label{eq:first1}
S_{cl}^1(v_1,v_2)=-2 \pi \mathrm{i} ( m^T\cdot \Omega \cdot m+n^T\cdot \Omega \cdot n),
\end{equation}
where $m\equiv\{m_0^1,m_1^1,m_0^2,m_1^2\}\in \mathbb{Z}^4$, $n\equiv\{n_0^1,n_1^1,n_0^2,n_1^2\} \in \mathbb{Z}^4$ and $\Omega$ is a symmetry matrix 
\begin{equation}
\Omega=\frac{ \mathrm{i} R^2}{4}
\begin{pmatrix}
A & -A & 2B & -2B  \\
-A & A & -2B & 2B \\
2B & -2B & 4C & -4C \\
-2B & 2B & -4C & 4C
\end{pmatrix}.
\end{equation}
Similarly, the second part of classical action can be given by 
\begin{equation}
\label{eq:second1}
S_{cl}^2(v_3,v_4)=-2 \pi \mathrm{i}\left(  {m^\prime}^T\cdot \Omega^\prime \cdot m^\prime+{n^\prime}^T\cdot \Omega^\prime \cdot n^\prime\right),
\end{equation}
where $m^\prime \equiv\{m_0^3,m_1^3,m_0^4,m_1^4\}\in \mathbb{Z}^4$, $n^\prime \equiv\{n_0^3,n_1^3,n_0^4,n_1^4\}\in \mathbb{Z}^4$ and
\begin{equation}
\Omega^\prime=\frac{ \mathrm{i} R^2}{ 4}
\begin{pmatrix}
A^\prime  & -A^\prime  &2 B^\prime  & -2B^\prime  \\
-A^\prime & A^\prime  & -2B^\prime & 2B^\prime \\ 
2B^\prime  & -2B^\prime &4 C^\prime   &  -4C^\prime  \\
-2B^\prime & 2B^\prime  & - 4C^\prime & 4C^\prime 
\end{pmatrix}.
\end{equation}

In the case of $N=2$ and $k=0$, there are no branch-point twist fields. Thus the classical summation is the 
same as \eqref{eq:third}. Putting \eqref{eq:first1}, \eqref{eq:second1} and \eqref{eq:third} all together, the instanton contribution of the partition function now is 
\begin{equation}
\label{eq:instantonsum2}
Z_{cl}=\left(\sum_{m \in \mathbb{Z}^4} e^{2 \pi \mathrm{i}  m^T\cdot \Upsilon \cdot m}\right)^2 \left(\sum_{m^\prime \in \mathbb{Z}^4} e^{2 \pi \mathrm{i}  {m^\prime}^T\cdot \Upsilon^\prime \cdot m^\prime}\right)^2,
\end{equation}
where 
\begin{equation}
\Upsilon=\frac{ \mathrm{i} R^2}{4}
\begin{pmatrix}
A+\beta & -A+\beta & 2B & -2B  \\
-A+\beta & A+\beta & -2B & 2B \\
2B & -2B & 4C & -4C \\
-2B & 2B & -4C & 4C
\end{pmatrix},\,
\Upsilon^\prime=\frac{ \mathrm{i} R^2}{4}
\begin{pmatrix}
A^\prime+\frac{1}{\beta}  & -A^\prime+\frac{1}{\beta}  &2 B^\prime  & -2B^\prime  \\
-A^\prime+\frac{1}{\beta} & A^\prime+\frac{1}{\beta}  & -2B^\prime & 2B^\prime \\ 
2B^\prime  & -2B^\prime &4 C^\prime   &  -4C^\prime  \\
-2B^\prime & 2B^\prime  & - 4C^\prime & 4C^\prime 
\end{pmatrix}
\end{equation}
Clearly the third and the fourth row of $\Upsilon(\Upsilon^\prime)$ are not independent. i.e, they are degenerate. By introducing a
matrix $U$
\begin{equation}
U=\begin{pmatrix}
1 & 0 & 0 & 0  \\
0 & 1 & 0 & 0\\
0 & 0 & 1 & 0 \\
0 & 0& 1 & 1
\end{pmatrix}, 
\end{equation} 
one can show that 
\begin{equation}
U\cdot \Upsilon \cdot U^T= \frac{ \mathrm{i} R^2}{4}
\begin{pmatrix}
A+\beta & -A+\beta & 2B & 0  \\
-A+\beta & A+\beta & -2B & 0 \\
2B & -2B & 4C & 0 \\
0 & 0 & 0 & 0
\end{pmatrix}.
\end{equation}
Now we can introduce a regulator $\epsilon>0$ as in \cite{eeoftwo}, and define
\begin{equation}
 U\cdot \Upsilon_\epsilon \cdot U^T= \frac{ \mathrm{i} R^2}{4}
\begin{pmatrix}
A+\beta & -A+\beta & 2B & 0  \\
-A+\beta & A+\beta & -2B & 0 \\
2B & -2B & 4C & 0 \\
0 & 0 & 0 & \epsilon
\end{pmatrix}=\Omega_\epsilon.
\end{equation}
We have the same relation for $\Upsilon^\prime$.
Since the upper left $3\times 3$ block of $\Omega_\epsilon$ is a Riemann matrix and $U$ is invertible, by using the identity of Riemann-Siegel theta function \cite{NIST}
\begin{equation}
\sum_{m \in \mathbb{Z}^4} e^{2 \pi \mathrm{i}  m^T\cdot \Upsilon_\epsilon \cdot m}=\Theta(0|2\Upsilon_\epsilon)=\Theta(0| U\cdot 2\Upsilon_\epsilon \cdot U^T)=\Theta(0|2\Omega_\epsilon),
\end{equation}
one can easily see that 
\begin{equation}\label{eq:cardymethod}
\begin{split}
Z_{cl}&= \lim_{\epsilon\rightarrow 0}\left(\sum_{m \in \mathbb{Z}^4} e^{2 \pi \mathrm{i}  m^T\cdot \Upsilon_\epsilon \cdot m}\right)^2 \left(\sum_{m^\prime \in \mathbb{Z}^4} e^{2 \pi \mathrm{i}  {m^\prime}^T\cdot \Upsilon_\epsilon^\prime \cdot m^\prime}\right)^2\\
&= \lim_{\epsilon\rightarrow 0}\left(\sum_{m \in \mathbb{Z}^4} e^{2 \pi \mathrm{i}  m^T\cdot \Omega_\epsilon \cdot m}\right)^2 \left(\sum_{m^\prime \in \mathbb{Z}^4} e^{2 \pi \mathrm{i}  {m^\prime}^T\cdot \Omega_\epsilon^\prime \cdot m^\prime}\right)^2.
\end{split}
\end{equation}
We then divide the vector of integer $m$ into $m=\tilde{m}+K$, where $K$ is eigenvector of $\Omega_\epsilon(
\Omega_\epsilon^\prime)$ with eigenvalue $\epsilon$ and $\tilde{m}$ is orthogonal to $K$.
As a result, we separate out the zero mode contribution which can be absorbed into the normalization.
Thus one can see that \eqref{eq:cardymethod} is equal to \eqref{eq:classical} up to the normalization.

\end{document}